{}
{}

\documentclass[11pt,a4paper]{article}

\newif\ifpublic\publictrue
\newif\iffancy\fancytrue

\usepackage[a4paper,text={160mm,247mm},centering]{geometry}
\usepackage{setspace}
\usepackage{amsmath,amssymb, amsfonts, geometry}
\makeatletter
\providecommand*{\shuffle}{%
  \mathbin{\mathpalette\shuffle@{}}%
}
\newcommand*{\shuffle@}[2]{%
  \sbox0{$#1\vcenter{}$}%
  \kern .15\ht0 
  \rlap{\vrule height .25\ht0 depth 0pt width 2.5\ht0}%
  \raise.1\ht0\hbox to 2.5\ht0{%
    \vrule height 1.75\ht0 depth -.1\ht0 width .17\ht0 %
    \hfill
    \vrule height 1.75\ht0 depth -.1\ht0 width .17\ht0 %
    \hfill
    \vrule height 1.75\ht0 depth -.1\ht0 width .17\ht0 %
  }%
  \kern .15\ht0 
}
\makeatother

\usepackage[pdfencoding=auto,bookmarks=true,hyperfigures=true]{hyperref}
\usepackage{graphicx}
\usepackage{float}
\usepackage[nosort]{cite}
\usepackage{lmodern}
\usepackage{amsbsy}
\usepackage[utf8]{inputenc}
\usepackage[font=small,labelfont=bf]{caption}
\usepackage{diagbox}
\usepackage{array}

\usepackage[usenames,dvipsnames]{xcolor}
\definecolor{dgreen}{rgb}{0,0.70,0.30}
\definecolor{gold}{rgb}{0.85,.66,0}
\definecolor{purple}{rgb}{1.0,0.3,0.6}

\usepackage{tikz}
\usetikzlibrary{calc} 
\usetikzlibrary{patterns} 
\usetikzlibrary{decorations.pathreplacing} 
\usetikzlibrary{decorations.markings} 
\usetikzlibrary{decorations.pathmorphing} 
\usetikzlibrary{positioning}

\usepackage{mpostinl}

\makeatletter
\newsavebox{\apb@box}\newlength{\apb@width}
\newcommand{\autoparbox}[2][c]{\sbox{\apb@box}{#2}%
 \settowidth{\apb@width}{\usebox{\apb@box}}%
 \parbox[#1]{\apb@width}{\usebox{\apb@box}}}

\newcommand{\includegraphicsboxex}[2][]{\autoparbox{\includegraphicsex[#1]{#2}}}
\makeatother

\iffancy

\def\showkeysrefformat#1{{\normalfont\tiny\ttfamily#1}}
\makeatletter
\def\SK@@ref#1>#2\SK@{%
 {\@inlabelfalse\leavevmode\vbox to\z@{%
 \vss\SK@refcolor\rlap{\vrule\raise .75em%
  \hbox{\showkeysrefformat{#2}}}}}}
\makeatother
\fi

\numberwithin{equation}{section}

\providecommand{\href}[2]{#2}

\makeatletter
\def\mr@ignsp#1 {\ifx\:#1\@empty\else #1\expandafter\mr@ignsp\fi}%
\newcommand{\multiref}[1]{\begingroup
\xdef\mr@no@sparg{\expandafter\mr@ignsp#1 \: }%
\def\mr@comma{}%
\@for\mr@refs:=\mr@no@sparg\do{\mr@comma\def\mr@comma{,}\ref{\mr@refs}}%
\endgroup}
\renewcommand{\eqref}[1]{(\multiref{#1})}
\makeatother

\makeatletter
\newcommand{\namedref}[2]{\hyperref[#2]{#1~\ref*{#2}}}
\newcommand{\secref}{\@ifstar{\namedref{Section}}{\namedref{section}}}
\newcommand{\subsecref}{\@ifstar{\namedref{Subsection}}{\namedref{subsection}}}
\newcommand{\appref}{\@ifstar{\namedref{Appendix}}{\namedref{appendix}}}
\newcommand{\tabref}{\@ifstar{\namedref{Table}}{\namedref{table}}}
\newcommand{\figref}{\@ifstar{\namedref{Figure}}{\namedref{figure}}}
\makeatother

\newcommand{\eqn}[1]{eq.~\eqref{#1}}

\newcommand{\eqns}[2]{eqs.~\eqref{#1} and~\eqref{#2}}

\newcommand{\rcite}[1]{ref.~\cite{#1}}

\providecommand{\hypersetup}[1]{}

\hypersetup{plainpages=false}
\hypersetup{pdfpagemode=UseNone}
\hypersetup{bookmarksnumbered=true}
\hypersetup{pdfstartview=FitH}
\hypersetup{colorlinks=false}
\hypersetup{citebordercolor={.5 1 .5}}
\hypersetup{urlbordercolor={.5 1 1}}
\hypersetup{linkbordercolor={1 .7 .7}}

\makeatletter
\let\@keywords\@empty
\let\@subject\@empty
\providecommand{\keywords}[1]{\gdef\@keywords{#1}}
\providecommand{\subject}[1]{\gdef\@subject{#1}}
\def\thetitle{\@title}
\def\theauthor{\@author}
\def\thesubject{\@subject}
\def\thedate{\@date}
\def\thekeywords{\@keywords}
\makeatother
\AtBeginDocument{
\hypersetup{pdftitle={\thetitle}}%
\hypersetup{pdfauthor={\theauthor}}%
\hypersetup{pdfsubject={\thesubject}}%
\hypersetup{pdfkeywords={\thekeywords}}%
}

\RequirePackage{verbatim}

\makeatletter
\newwrite\bibinl@out
\newenvironment{bibtex}[1][\jobname]{%
 \immediate\openout\bibinl@out #1.bib
 \immediate\write\bibinl@out{\@percentchar generated from `\jobname' starting line \the
\inputlineno^^J}%
 \def\verbatim@processline{\immediate\write\bibinl@out{\the\verbatim@line}}%
 \@bsphack\let\do\@makeother\dospecials\catcode`\^^M\active\verbatim@start
}%
{\immediate\closeout\bibinl@out\@esphack}
\makeatother

\newif\ifnote 
\notetrue

\allowdisplaybreaks

\newcommand{\ba}  {\begin{array}}
\newcommand{\ea}  {\end{array}}
\newcommand{\bdm} {\begin{displaymath}}
\newcommand{\edm} {\end{displaymath}}

\newcommand{\bea} {\begin{equation}\ba{lcl}}
\newcommand{\eea} {\ea\end{equation}}

\newcommand{\bc}  {\begin{center}}
\newcommand{\ec}  {\end{center}}

\def\beq{\begin{equation}}
\def\eeq{\end{equation}}

\newcommand{\vecb}{\left(\begin{array}{c}}
\newcommand{\vece}{\end{array}\right)}
\newcommand{\ccb}{\left(\begin{array}{cc}}
\newcommand{\cce}{\end{array}\right)}
\newcommand{\cccb}{\left(\begin{array}{ccc}}
\newcommand{\ccce}{\end{array}\right)}
\newcommand{\ccccb}{\left(\begin{array}{cccc}}
\newcommand{\cccce}{\end{array}\right)}
\newcommand{\cccccb}{\left(\begin{array}{ccccc}}
\newcommand{\ccccce}{\end{array}\right)}

\newcommand{\pd}{\partial}

\newcommand{\z}{\zeta}

\newcommand{\SL}{\mathrm{SL}}

\newcommand{\te}{\textrm}

\newcommand{\dd}{\mathrm{d}}

\newcommand{\ZZ}{\mathbb Z}

\newcommand{\CL}{\mathcal L}
\newcommand{\CN}{\mathcal N}
\newcommand{\CO}{\mathcal O}

\def\MHV{\mathrm{MHV}}

\def\reg{\textrm{reg}}
\def\MRK{\textrm{MRK}}
\def\BFKL{\textrm{BFKL}}

\newcommand{\nnl}{\nonumber\\}

\DeclareMathOperator{\zm}{\zeta}
\DeclareMathOperator{\Li}{Li}

\DeclareMathOperator{\omm}{\omega}

\newcommand{\oz}{1\!-\!z}
\newcommand{\oy}{1\!-\!y}
\newcommand{\ozb}{1\!-\!\bar{z}}

\newcommand{\zb}{\bar{z}}
\newcommand{\yb}{\bar{y}}

\newcommand{\tblue}[1]{\textcolor{blue}{#1}}

\usepackage{amsthm} 
\theoremstyle{plain}

\vspace*{1.8cm}
\title{\textbf{Towards single-valued polylogarithms in two variables\\
for the seven-point remainder function\\
in multi-Regge-kinematics}}
\author{
Johannes Broedel$^{\te{a}}$, 
Martin Sprenger$^{\te{b}}$,
Alejandro Torres Orjuela$^{\te{a,c}}$
}
\date{\today}

\begin{document}

\input{MRK7_fig.fig}

\pdfbookmark[1]{Title Page}{title} \thispagestyle{empty}
\begin{flushright}
  \verb!HU-EP-16/19!\\
  \verb!HU-Mathematik-2016-11!
\end{flushright}
\vspace*{0.4cm}

\begin{center}%
  \begingroup\LARGE\bfseries\thetitle\par\endgroup
  \vspace{1.1cm}

\begingroup\large\theauthor\par\endgroup
\vspace{8mm}%

\begingroup\itshape
$^{\te{a}}$Institut f\"ur Mathematik und Institut f\"ur Physik,\\
Humboldt-Universit\"at zu Berlin\\
IRIS Adlershof, Zum Gro\ss{}en Windkanal 6, 12489 Berlin, Germany
\par\endgroup
\vspace{4mm}
\begingroup\itshape
$^{\te{b}}$Institut f\"ur Theoretische Physik,\\
Eidgen\"ossische Technische Hochschule Z\"urich\\
Wolfgang-Pauli-Strasse 27, 8093 Z\"urich, Switzerland
\par\endgroup
\vspace{4mm}
\begingroup\itshape
$^{\te{c}}$Institut f\"ur Mathematik,\\
Technische Universit\"at Berlin\\
Stra\ss e des 17.~Juni 136, 10623 Berlin, Germany
\par\endgroup

\vspace{0.9cm}

\begingroup\ttfamily
jbroedel@physik.hu-berlin.de\\
sprengerm@itp.phys.ethz.ch\\
alejandro.torresorjuela@campus.tu-berlin.de
\par\endgroup

\vspace{1.2cm}

\bigskip

\textbf{Abstract}\vspace{5mm}

\begin{minipage}{13.4cm}
We investigate single-valued polylogarithms in two complex variables, which are
relevant for the seven-point remainder function in $\CN=4$ super-Yang--Mills
theory in the multi-Regge regime. After constructing these two-dimensional
polylogarithms, we determine the leading logarithmic approximation of the
seven-point remainder function up to and including five loops. 
\end{minipage}

\vspace*{4cm}

\end{center}

\newpage

\setcounter{tocdepth}{2}
\tableofcontents


\section{Introduction}
\label{sec:introduction}

One of the most interesting results in the study of scattering amplitudes in
planar $\CN=4$ super-Yang--Mills theory is that certain amplitudes can be
constructed to high loop orders solely from understanding the space of
functions describing the amplitude, its symmetries as well as its limiting
behavior in special kinematic regimes. Indeed, following this program of
bootstrapping the six-point amplitude, the remainder function is by now known
up to four loops \cite{Dixon:2013eka,Dixon:2014voa,Dixon:2014iba,Dixon:2015iva}
and it is very likely that this program can be continued to higher orders.

Much less, however, is known about the seven-point amplitude.  The MHV
remainder function in general kinematics has been calculated up to two loops
\cite{Golden:2014xqf}, while the symbol is known up to three loops
\cite{CaronHuot:2011ky, Drummond:2014ffa}.
It is therefore a sensible idea to consider specific kinematic configurations,
hoping to obtain higher-order results in those special settings which can then
be used as constraints on a potential ansatz for the full seven-point remainder
function.

The special kinematic configuration considered in this paper is the multi-Regge
limit.  This limit has been studied in the seven-point case before: the
Mandelstam regions have been classified \cite{Bartels:2013jna,
Bartels:2014jya}, and the remainder function has been calculated in the most
interesting Mandelstam region in the leading logarithmic approximation (LLA)
\cite{Bartels:2011ge}. Furthermore, the seven-point remainder function has been
studied at strong coupling \cite{Bartels:2014ppa, Bartels:2014mka} as well as
from the perspective of the symbol for two \cite{Bargheer:2015djt} and three
loops \cite{Bargheer:2016eyp}.

In this paper, we follow the path of understanding for the six-point remainder
function in the multi-Regge limit: while first expressed in Fourier--Mellin
space up to next-to-leading logarithmic order (NLLA) \cite{Bartels:2008ce,
Bartels:2008sc, Fadin:2011we}, the identification of the relevant space of
functions paved the way for an efficient evaluation of the remainder function
in momentum space \cite{Dixon:2012yy, Drummond:2015jea, Broedel:2015nfp}. In
the six-point case, these functions are single-valued harmonic polylogarithms
(SVHPLs) which we will briefly review below. 

Similarly, in \rcite{Bartels:2011ge}, the seven-point MHV remainder function
was written down in Fourier--Mellin space. Due to the complicated nature of the
integral, it was so far only evaluated up to two loops. In this paper we
therefore set out to identify a suitable two-variable generalization of SVHPLs
constituting the relevant space of functions in the seven-point case. We
construct those functions from their differential behavior, which allows us to
obtain results up to five loops.
   
The paper is organized as follows. In \secref{sec:6pt} we review the six-point
case and the construction of the SVHPLs describing the remainder function. In
\subsecref{ssec:7ptbasics} we then move on to seven gluons and highlight the
differences to the six-point case and the two-dimensional generalization of
HPLs in \subsecref{ssec:2dHPL} before constructing the two-dimensional analogue of
SVHPLs in \subsecref{ssec:2dsvHPL}.
Using these functions, we obtain expressions for the remainder function in LLA
up to five loops in \subsecref{ssec:results}, before concluding in
\secref{sec:conclusion}. 


\section{Six-point remainder function in multi-Regge kinematics}
\label{sec:6pt}

\subsection{Starting point/setup}
\label{ssec:starting6}

The six-point remainder function $R_6^\MHV$ in $\mathcal{N}=4$
super-Yang--Mills theory\footnote{For simplicity of presentation, we confine
  ourselves to the MHV remainder function; the six-point remainder function
  for NMHV is described and evaluated in
refs.~\cite{Dixon:2012yy,Dixon:2014iba,Dixon:2015iva,Broedel:2015nfp}.}
describes the discrepancy between the full amplitude and the BDS ansatz
\cite{Bern:2005iz}, 
\begin{align}
  A^\MHV_6&=A_{\text{BDS}}\,e^{R_6^\MHV}\label{eqn:BDSR}\,.
\end{align}
While the calculation of the six-point remainder function in general kinematics
requires a multitude of different techniques, its determination in the
so-called multi-Regge kinematics is simpler.  The multi-Regge limit
refers to the kinematical regime of the scattering amplitude in which the
rapidities of the outgoing particles are strongly ordered.  In terms of
external momenta $k_i$, the multi-Regge limit can be most easily described
using dual variables $x_{ii+1}:=x_i-x_{i+1}:=k_i$. Expressed in terms of dual
conformal cross ratios $u_i$, the multi-Regge behavior reads:
\begin{equation}
	u_1:=\frac{x_{13}^2x_{46}^2}{x_{14}^2x_{36}^2}\rightarrow 1, \quad u_2:=\frac{x_{24}^2x_{15}^2}{x_{25}^2x_{14}^2}\rightarrow 0, \quad u_3:= \frac{x_{35}^2x_{26}^2}{x_{36}^2x_{25}^2}\rightarrow 0,
	\label{eq:mrl_crs6}
\end{equation}
where the reduced cross ratios
\begin{equation}
	\tilde{u}_2:=\frac{u_2}{1-u_1}=:\frac{1}{|1+w|^2},\quad \tilde{u}_3:=\frac{u_3}{1-u_1}=:\frac{|w|^2}{|1+w|^2}
	\label{eq:mrl_red_crs6}
\end{equation}
are kept finite. As visible from \eqn{eq:mrl_red_crs6}, six-point multi-Regge
kinematics is completely determined by a complex parameter $w$ and the large
cross ratio $u_1$.  

While the na\"ive limit \eqn{eq:mrl_crs6} yields a vanishing remainder
function, a non-trivial result can be obtained by an analytic continuation to
the so-called Mandelstam region \cite{Bartels:2008ce, Bartels:2008sc}, which is
implemented by a clockwise continuation of the large cross ratio $u_1$,
\begin{equation}
	u_1\rightarrow e^{-2i\pi}u_1
\end{equation}
before taking the limit (\ref{eq:mrl_crs6}).  For this setup, the six-point
remainder function in multi-Regge kinematics can be written as a
Fourier--Mellin integral \cite{Bartels:2008sc, Lipatov:2010ad, Fadin:2011we}:
\begin{align}
  e^{R_6^\MHV+i\pi\delta}|_\MRK=&\cos \pi\omm_{ab}+i\frac{a}{2}\sum_{n=-\infty}^{\infty}(-1)^n
  \left(\frac{w}{w^*}\right)^{\frac{n}{2}}\int_{-\infty}^\infty\frac{\dd\nu}{\nu^2+\frac{n^2}{4}}\,|w|^{2i\nu}\Phi_\reg^\MHV(\nu,n)\nnl
  &\qquad\qquad\qquad\times\,\exp\left[-\omm(\nu,n)\left(\log(1-u_1)+i\pi+\frac{1}{2}\log\frac{|w|^2}{|1+w|^4}\right)\right]\,.
  \label{eqn:master6}
\end{align}
In the above equation, the first term originates from a Regge pole exchange, while
the second term comes from the exchange of a two-Reggeon bound state, which gives
rise to a Regge-cut contribution. 
The so-called impact factor $\Phi^\MHV_\reg(\nu,n)$ and the BFKL eigenvalue $\omm(\nu,n)$
appearing in the latter have an expansion in powers of the loop-counting
parameter $a=\frac{g^2 N_c}{8 \pi^2}$:
\begin{align}
	\omm(\nu,n)&=-a(E_{\nu,n}+aE^{(1)}_{\nu,n}+a^2E^{(2)}_{\nu,n})+\CO(a^4),\nnl
  \Phi_\reg^\MHV&=1+a\,\Phi^{(1),\MHV}+a^2\,\Phi^{(2),\MHV}+\CO(a^3)\,.
  \label{eqn:BFKLloop6}
\end{align}
Physically, the BFKL eigenvalue $\omega(\nu,n)$ describes the evolution of the
two-Reggeon bound state, while the impact factor describes the coupling of the
two-Reggeon bound state to the physical gluons. While the first orders of these
quantities were determined by direct calculation \cite{Bartels:2008sc,
Lipatov:2010ad, Fadin:2011we, Dixon:2012yy, Dixon:2014voa}, a general solution
based on the Wilson-loop OPE \cite{Basso:2013vsa, Basso:2013aha, Basso:2014koa,
Basso:2014nra} has been identified in \rcite{Basso:2014pla}.

In the six-point (and seven-point) calculations below, we will only need the
lowest-order term of the BFKL eigenvalue, which reads
\begin{equation}
	E_{\nu,n}=-\frac{1}{2}\frac{|n|}{\nu^2+\frac{n^2}{4}}+\psi\left(1+i\nu+\frac{|n|}{2}\right)+\psi\left(1-i\nu+\frac{|n|}{2}\right)-2\psi(1),
	\label{eq:e_lla}
\end{equation}
where $\psi(x)$ is the digamma function.  The two other quantities appearing in
\eqn{eqn:master6}, the phase $\delta$ and the Regge-pole contribution
$\omm_{ab}$, are related to the cusp anomalous dimension $\gamma_K(a)$ via
\begin{align}
  \omm_{ab}=\frac{1}{8}\gamma_K(a)\log|w|^2\quad \text{and} \quad 
  \delta=\frac{1}{8}\gamma_K(a)\log\frac{|w|^2}{|1+w|^4}\,
\end{align}
and are thus known to all orders \cite{Beisert:2006ez}.

Finally, the term $\log(1-u_1)$ in the integrand of \eqn{eqn:master6} is large
because of the behavior of the cross ratio $u_1$ in the multi-Regge limit
\eqn{eq:mrl_crs6}. This suggest to organize the remainder function in powers of
$\log(1-u_1)$ at each loop order:
\begin{equation}
  R_6^\MHV\big|_\MRK=2\pi i\sum\limits_{\ell=2}^\infty\sum\limits_{n=0}^{\ell-1}a^\ell \log^n(1-u_1)\Big[g^{(\ell)}_n(w,w^*) + 2 \pi i \,h^{(\ell)}_n(w,w^*)\Big]\,.
  \label{eq:rem_exp_gh6}
\end{equation}
In the above equation, all terms with $n\!=\!\ell\!-\!1$ are referred to as the
leading logarithmic approximation (LLA) and terms with $n\!=\!\ell\!-\!1\!-\!k$
belong to (Next-to)$^k$-LLA. Since the imaginary and real parts $g^{(\ell)}_n$
and $h^{(\ell)}_n$ are not independent \cite{Dixon:2012yy}, it is sufficient to
calculate all imaginary parts $g^{(\ell)}_n$ in order to determine the full
remainder function.  For more details on the six-point remainder function in
the multi-Regge limit we refer the reader to \rcite{Broedel:2015nfp} and
continue with the description of the relevant functions for the evaluation of
\eqn{eqn:master6}.


\subsection{Single-valued harmonic polylogarithms in one variable}
\label{ssec:1dsvhpl}

Before describing the functions governing the integral \eqn{eqn:master6}, let
us introduce harmonic polylogarithms (or HPLs for short) \cite{Remiddi:1999ew},
which are defined as iterated integrals
\begin{equation}
  \label{eqn:polylogdef}
  H_{a_1, a_2, \ldots, a_n}(z)=\int_0^z \dd t f_{a_1}H_{a_2, \ldots, a_n}(t)\,,
\end{equation}
where the integration weights $f_a$ are given as 
\begin{equation}
  \label{eqn:intweight}
  f_{-1}=\frac{1}{1+t},\quad f_{0}=\frac{1}{t},\quad\textrm{and}\quad 
  f_{1}=\frac{1}{1-t} \,.
\end{equation}
In \eqn{eqn:polylogdef}, the length of the index vector $\vec{a}=\lbrace
a_1,\ldots,a_n\rbrace$ is called the \textit{weight} of a HPL, while $z$ is
referred to as the \textit{argument}.  For the six-point remainder function,
only the latter two integration weights in \eqn{eqn:intweight} will appear.
Corresponding to the weights $f_{0}$ and $f_{1}$ we introduce two letters $x_0$
and $x_1$ which will be used as non-commutative bookkeeping variables in
generating functions for polylogarithms below.

From their definition \eqn{eqn:polylogdef} it is clear that HPLs satisfy the 
differential equation
\begin{equation}
	\frac{\pd}{\pd z}H_{a_1, a_2, \ldots, a_n}= f_{a_1}(z)H_{a_2, \ldots, a_n}(z).
	\label{eqn:diffhpl}
\end{equation}
Harmonic polylogarithms of low weight can be conveniently
expressed in terms of logarithms and dilogarithms, for example
\begin{align}
  H_{\underbrace{\scriptstyle{0, \ldots, 0}}_w}(z)=\frac{1}{w!}\log^w(z),\quad
  H_{\underbrace{\scriptstyle{1, \ldots, 1}}_w}(z)=\frac{1}{w!}(-\log(1-z))^w,\quad
  H_{\underbrace{\scriptstyle{0, \ldots, 0}}_{(w-1)},1}(z)&=\Li_w(z)\,.
\end{align}
Furthermore, harmonic polylogarithms satisfy a scaling identity
\begin{equation}
  \label{eqn:scaling}
  H_{k\cdot a_1,\ldots, k\cdot a_n}(k\cdot z)=H_{a_1,\ldots, a_n}(z)\quad \textrm{for}\quad k\neq0,\,z\neq 0\,,
\end{equation}
which is valid whenever $a_n\neq 0$. 

Given that the usual logarithm has a branch cut which is canonically chosen to
lie along the negative real axis, its iterated and integrated versions have
branch cuts, as well. However, it is possible to obtain single-valued harmonic
polylogarithms (SVHPLs) by linearly combining products of the form
$H_{s_1}(z)H_{s_2}(\zb)$ in a way that all branch cuts cancel. The
combinations of HPLs leading to SVHPLs are unique and can be determined from
demanding triviality of the monodromies around singular points of HPLs
\cite{Brown2004527}.  

The lowest-weight SVHPLs\footnote{As will be clear from the examples given in
\eqn{eqn:svhpl}, the $\CL_s(z)$ are functions of both $z$ and $\zb$. For
simplicity, however, we will denote these functions as $\CL_s(z)$.} read
\begin{align}
  \CL_0(z)&=H_0(z)+H_0(\zb)\nnl
  \CL_1(z)&=H_1(z)+H_1(\zb)\nnl
  \CL_{0,0}(z)&=H_{0,0}(z)+H_{0,0}(\zb)+H_0(z)H_0(\zb)\nnl
  \CL_{1,0}(z)&=H_{1,0}(z)+H_{0,1}(\zb)+H_1(z)H_0(\zb)\nnl
  \CL_{1,0,1}(z)&=H_{1,0,1}(z)+H_{1,0,1}(\zb)+H_{1,0}(z)H_1(\zb)+H_{1}(z)H_{1,0}(\zb)\nnl
  &\quad\vdots
  \label{eqn:svhpl}
\end{align}
While up to weight three the expressions follow an obvious pattern,
$\zm$-values make an appearance starting at weight four, for example 
\begin{align}
  \CL_{1,0,1,0}(z)&=H_{1,0,1,0}(z)+H_{0,1,0,1}(\zb)+H_{1,0,1}(z)H_{0}(\zb)+H_{1}(z)H_{0,1,0}(\zb)\nnl
	       &\qquad+H_{1,0}(z)H_{01}(\zb)-4\zm_3H_1(\zb).
	       \label{eq:ex_zeta_svhpl}
\end{align}
An elaborate introduction to SVHPLs in the context of the six-point remainder
function in MRK in which the method for solving the single-valuedness condition is
carefully explained can be found in section 3 of \rcite{Dixon:2012yy}.

In the remainder of this subsection, let us collect several properties of
SVHPLs which will be useful below: Two SVHPLs labeled by words $s_1$ and $s_2$
satisfy the shuffle relation
\begin{equation}
  \CL_{s_1}(z)\,\CL_{s_2}(z)=\sum\limits_{s\in s_1\shuffle s_2}\CL_s(z)\,,
\end{equation}
where the shuffle $s_1\shuffle s_2$ refers to all permutations of $s_1 \cup s_2$ which
leave the order of elements in $s_1$ and $s_2$ unaltered.  
The generating functional for the SVHPLs,
\begin{equation}
  \CL^{\lbrace 0,1\rbrace}(z)=\sum\limits_{s\in X(\lbrace x_0,x_1\rbrace)}\CL_s(z)s=1+\CL_0(z)\,x_0+\CL_1(z)\,x_1+\CL_{0,0}(z)\,x_0x_0+\CL_{0,1}(z)\,x_0x_1+\dots\,,
	\label{eq:genfunc}
\end{equation}
where $X(\lbrace x_0,x_1\rbrace)$ are all words\footnote{For convenience,
SVHPLs will be labeled by the indices of the letters rather than by the
letters themselves.} in the alphabet $\lbrace x_0,x_1\rbrace$. 
This generating functional satisfies the differential equations
\begin{equation}
  \frac{\pd}{\pd z}\CL^{\lbrace 0,1\rbrace}(z)=\left(\frac{x_0}{z}+\frac{x_1}{1-z}\right)\CL^{\lbrace 0,1\rbrace}(z),\quad 
  \frac{\pd}{\pd \zb}\CL^{\lbrace 0,1\rbrace}(z)=\CL^{\lbrace 0,1\rbrace}(z)\left(\frac{y_0}{\zb}+\frac{y_1}{1-\zb}\right)\,,
	\label{eqn:svhpldiff}
\end{equation}
where $\lbrace y_0,y_1\rbrace$ is an additional alphabet, which appears in the
construction of SVHPLs in \rcite{Brown2004527} and is related by the single-valuedness
condition to the alphabet $\lbrace x_0,x_1\rbrace$ mentioned above. 
Solving this condition order by order, one finds
\begin{align}
  y_0&=x_0\quad\text{and}\nnl
  y_1&=x_1-\z_{3}(2x_0x_0x_1x_1-4x_0x_1x_0x_1+2x_0x_1x_1x_1+4x_1x_0x_1x_0+\cdots)+\cdots\,,
  \label{eqn:fixpointsol}
\end{align}
and can thus find the analogue of \eqn{eqn:svhpl} for $\CL_s$ for an arbitrary
label $s$ constructed from the alphabet $\lbrace x_0,x_1 \rbrace$.  Note,
however, that from \eqn{eqn:svhpldiff} the $\CL_s(z)$ satisfy the same
differential equation in $z$ as the corresponding HPL $H_s(z)$, cf.
\eqn{eqn:diffhpl}.

\subsection{Calculation of the six-point remainder function}
As pointed out at the end of \subsecref{ssec:starting6}, the problem of
calculating the remainder function $R^\MHV_6|_\MRK$ via \eqn{eqn:master6} boils
down to evaluating the real part of the sum over the integral, which will yield
the functions $g^{(\ell)}_n$. The crucial ingredients here are the loop
expansions of the impact factor $\Phi^\MHV_\reg(\nu,n)$ and the BFKL eigenvalue
$\omm(\nu,n)$ in \eqn{eqn:BFKLloop6}.  

To calculate $g^{(\ell)}_n$ one would then expand the integral to the desired loop
and logarithmic order, close the contour at infinity and sum up the residues.
This, however, becomes cumbersome already beyond the lowest loop order.

As discussed in \rcite{Dixon:2012yy}, the functions $g_n^{(\ell)}$ can be
expressed in terms of SVHPLs. This opens a natural and simpler way for the
calculation of \eqn{eqn:master6}: one starts from an ansatz in SVHPLs and
compares the series expansions in $(w,w^\ast)$ of both the ansatz and the
integral.  Following this approach  the remainder function was calculated
up to five loops, as well as for higher loop orders in
LLA and NLLA \cite{Dixon:2012yy,Dixon:2014voa,Drummond:2015jea}.

A more direct evaluation of the remainder function \eqn{eqn:master6} was
developed in \rcite{Broedel:2015nfp}. The key insight, first used in
refs.~\cite{Drummond:2013nda,Drummond:2015jea}, is that the leading term of any
SVHPL is simply given by the harmonic polylogarithm with the same index
structure
\begin{equation}
        \CL_s(w,w^\ast)=H_s(w)+\dots,
        \label{eq:svhpl_lc}
\end{equation}
as exemplified in \eqn{eqn:svhpl}. Importantly, the term $H_w(w)$ is the only
term in the expansion of the SVHPL which does not depend on $w^\ast$.
Comparing with the dispersion relation \eqref{eqn:master6}, we see that the
leading terms are encoded in the residues at $\nu=-\frac{i n}{2}$ as only for
those poles the residues will have no contribution from $w^\ast$.

Since the remainder function can be expressed in terms of SVHPLs exclusively, a
viable approach consists of simply determining the leading terms, which will be
a linear combination of HPLs and obtaining the full result by simply promoting
HPLs to SVHPLs via
\begin{equation}
        H_s(w)\rightarrow \CL_s(w,w^\ast)\,.
        \label{eq:replacement_svhpls}
\end{equation}
In performing the above replacement, the contributions from the omitted
residues are restored automatically. 

In \rcite{Broedel:2015nfp}, we started from this observation and identified
recursion relations between different integrals which hold on the locus of the
poles $\nu=-\frac{i n}{2}$, but which by using \eqn{eq:replacement_svhpls} lift
to relations of the full result. Employing these relations, we reduced all
integrals to a set of trivial basis integrals. This allowed us to efficiently
evaluate the remainder function up to very high loop- and logarithmic orders
and to prove Pennington's formula \cite{Pennington:2012zj} for the six-point
remainder function in LLA.

Given the success of this approach for the six-point remainder function, it is
natural to ask, whether a similar formalism can be established for seven
points. This idea is going to be discussed in the next section. 

\section{Seven-point remainder function in multi-Regge kinematics}
\label{sec:7pt}

\subsection{From six to seven gluons}
\label{ssec:7ptbasics}
\begin{figure}
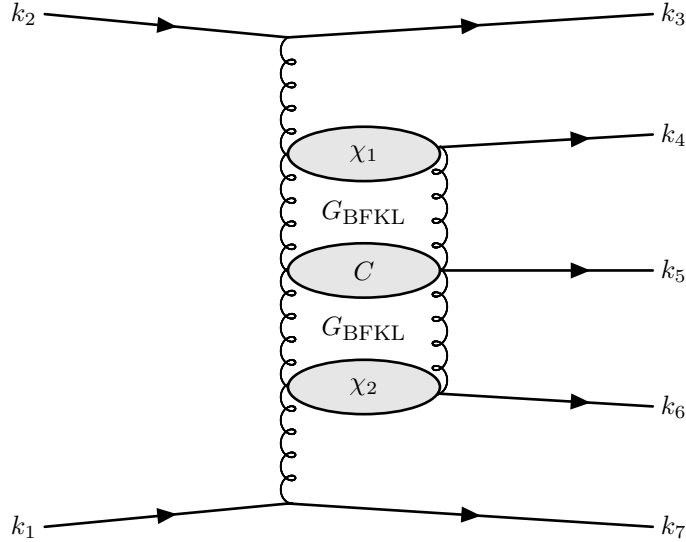

  \begin{center}
\includegraphicsboxex{figure1.mps}
\end{center}
\caption{Pictorial representation of eqs.~(\ref{eq:master7}) and
(\ref{eq:r7_int}), with $C$ being the central emission vertex, $G_{\BFKL}$
describing the evolution of the two-Reggeon bound state and $\chi_1$, $\chi_2$
being the building blocks of the impact factor.}
\label{fig:r7_mmm}
\end{figure}

We now move on to the seven-point MHV remainder function in multi-Regge
kinematics, which is defined similarly to the six-point case,
\begin{equation}
	A_{7}^\MHV=A_{\mathrm{BDS}} e^{R_7^\MHV}.
	\label{eq:7pt_def_R}
\end{equation}
The kinematics in this case is governed by seven conformal cross ratios,
\begin{alignat}{3}
\nonumber u_{1,1}&=\frac{x_{37}^2x_{46}^2}{x_{47}^2x_{36}^2},\quad &&u_{2,1}=\frac{x_{15}^2x_{24}^2}{x_{14}^2x_{25}^2},\quad &&u_{3,1}=\frac{x_{35}^2x_{26}^2}{x_{25}^2x_{36}^2},\\
u_{1,2}&=\frac{x_{14}^2x_{57}^2}{x_{15}^2x_{47}^2},\quad &&u_{2,2}=\frac{x_{16}^2x_{25}^2}{x_{15}^2x_{26}^2},\quad &&u_{3,2}=\frac{x_{36}^2x_{27}^2}{x_{26}^2x_{37}^2},\label{eq:7pt_crs}\\
\nonumber \tilde{u}&=\frac{x_{13}^2x_{47}^2}{x_{37}^2x_{14}^2}. && && 
\end{alignat}
In the multi-Regge limit the cross ratios $u_{1,s}$, $s=1,2$, and $\tilde{u}$
approach $1$, while all other cross ratios tend to zero.  Due to a conformal
Gram relation, only six of the above cross ratios are independent.  It is,
however, advantageous for what follows to keep all seven cross ratios
explicitly.  The remaining kinematic freedom in the multi-Regge limit is again
given by the reduced cross ratios
\begin{equation}
	\tilde{u}_{a,s}:=\frac{u_{a,s}}{1-u_{1,s}},
	\label{eq:def_red_crs}
\end{equation}
which are finite in the multi-Regge limit and which we again parameterize by two
complex variables $w_1$, $w_2$ defined as
\begin{equation}
	\tilde{u}_{2,1}=:\frac{1}{|1+w_1|^2},\quad \tilde{u}_{3,1}=:\frac{|w_1|^2}{|1+w_1|^2},\quad
	\tilde{u}_{2,2}=:\frac{1}{|1+w_2|^2},\quad \tilde{u}_{3,2}=:\frac{|w_2|^2}{|1+w_2|^2}.
	\label{eq:def_ws}
\end{equation}
\begin{figure}
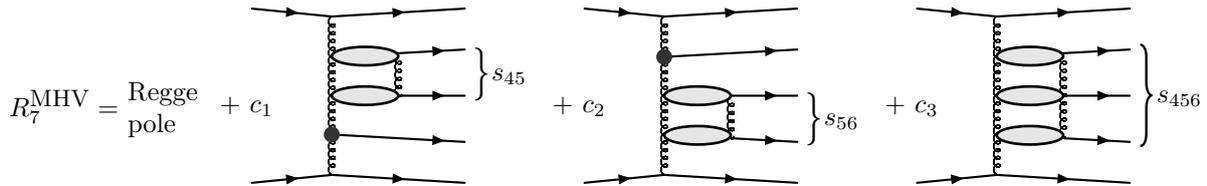

  \begin{center}
\includegraphicsboxex{figure2.mps}
\end{center}
\caption{In the seven-point case, the remainder function in every Regge region
can be written as a linear combination of the Regge pole contribution, the
\textit{short cuts} in the $s_{45}$- and $s_{56}$-channel and the \textit{long
cut} in the combined $s_{456}$-channel. See \cite{Bartels:2013jna,
Bartels:2014jya} for details.}
\label{fig:regge_regions}
\end{figure}
A key difference to the six-point case is that several interesting Mandelstam
regions exist, in which Regge cuts appear. However, as is shown in
\cite{Bartels:2013jna, Bartels:2014jya}, the seven-point remainder function can
be written as a linear combination of three elementary building blocks in every
Regge region.  These building blocks are usually called the \textit{short cuts}
which describe a Regge cut in the $s_{45}$- and $s_{56}$-channel, respectively, as
well as the \textit{long cut} which describes a Regge cut which spans the
$s_{456}$-channel, see \cite{Bartels:2013jna, Bartels:2014jya} for
details and figure \ref{fig:regge_regions} for a pictorial representation.
As it turns out, the short cuts are fully determined by the BFKL
eigenvalue and impact factor of the six-point amplitude. Therefore, we focus on
the long cut in which a new ingredient appears.  To study the long cut, one can
take the analytic continuation\footnote{Subtleties in the choice of path that
arise due to the conformal Gram relation are, for example, discussed in
\rcite{Bartels:2014mka}, but do not play a role here.}
\begin{equation}
	\tilde{u}\rightarrow e^{-2\pi i}\tilde{u}
	\label{eq:rot_p7mmm}
\end{equation}
of the remainder function, before going to the multi-Regge limit.  In this
Mandelstam region, the short cuts do not contribute and the remainder function
is fully determined by the long cut.  
In LLA, the remainder function in this
region was stated in \rcite{Bartels:2014jya} and reads
\begin{align}
	R_7^\MHV=1+i\pi\sum\limits_{\ell=2}^\infty\sum\limits_{k=0}^{\ell-1}\frac{a^\ell}{k!(\ell-1-k)!}\log^k(1-u_{1,1})\log^{\ell-1-k}(1-u_{1,2})\,\mathcal{I}_7\left[E_{\nu,n}^k\,E_{\mu,m}^{\ell-k-1}\right]\,,
	\label{eq:master7}
\end{align}
where we define
\begin{align}
	\mathcal{I}_7\left[\mathcal{F}(\nu,n,\mu,m)\right]:=\sum\limits_{n=-\infty}^\infty\sum\limits_{m=-\infty}^\infty\,\int\limits_{-\infty}^{+\infty}\frac{d\nu}{2\pi}\,\int\limits_{-\infty}^{+\infty}\frac{d\mu}{2\pi}\,w_1^{i\nu+n/2}&(w_1^\ast)^{i\nu-n/2} w_2^{i\mu+m/2}(w_2^\ast)^{i\mu-m/2}\label{eq:r7_int}\\
\nonumber	&\times C(\nu,n,\mu,m)\, \mathcal{F}(\nu,n,\mu,m),
\end{align}
and where
\begin{equation}
	C(\nu,n,\mu,m)=(-1)^{n+m}\frac{\Gamma\left(-i\nu-\frac{n}{2}\right)\Gamma\left(i\mu+\frac{m}{2}\right)\Gamma\left(i(\nu-\mu)+\frac{m-n}{2}\right)}
	{\Gamma\left(1+i\nu-\frac{n}{2}\right)\Gamma\left(1-i\mu+\frac{m}{2}\right)\Gamma\left(1-i(\nu-\mu)+\frac{m-n}{2}\right)}
	\label{eq:def_cev}
\end{equation}
is the so-called central emission vertex.  Comparing expression
(\ref{eq:r7_int}) with the corresponding equation for the six-point case
(\ref{eqn:master6}), we see that this is a new ingredient.
Like in the six-point case there is again a nice physical interpretation
of all the terms in \eqns{eq:r7_int}{eq:master7}, with the
central emission vertex $C(\nu,n,\mu,m)$ describing the emission of a physical gluon
from a bound state of two reggeized gluons and the BFKL eigenvalue $E_{\nu,n}$
describing the evolution of the bound state of two reggeized gluons,
of which we have two because of the appearance of the central
emission vertex.  The impact factor which describes the coupling of the bound
state to the physical gluons does not appear in \eqn{eq:r7_int} since it is
trivial in LLA.  We can represent \eqn{eq:master7}
graphically as shown in figure \ref{fig:r7_mmm}.  Note that \eqn{eq:master7}
has a similar form to \eqn{eq:rem_exp_gh6}, only with two distinct large
logarithms. Furthermore, as always in LLA, the real part vanishes.

Let us now study the symmetry properties of \eqn{eq:r7_int}. Using the expressions
for the LLA BFKL eigenvalue \eqn{eq:e_lla} and the central emission vertex
\eqn{eq:def_cev} we see that the remainder function is invariant under exchange
of
\begin{equation}
	\nu\leftrightarrow -\mu,\, n\leftrightarrow -m,
	\label{eq:sym_munu1}
\end{equation}
which corresponds to a swap
\begin{equation}
	w_1\leftrightarrow\frac{1}{w_2},
	\label{eq:sym_w1}
\end{equation}
as well as under exchange of
\begin{equation}
	n\leftrightarrow -n,\,m\leftrightarrow -m
	\label{eq:sym_munu2}
\end{equation}
which, in turn, corresponds to
\begin{equation}
	w_1\leftrightarrow w^\ast_1,\,w_2\leftrightarrow w^\ast_2.
	\label{eq:sym_w2}
\end{equation}
These symmetry properties will be very useful when evaluating the remainder
function later on.

To evaluate \eqn{eq:r7_int}, one first closes the contours of the two integrals
at infinity and then sums up the residues.  A convenient choice is to close the
contour of the $\mu$-integration in the upper half-plane and the contour of the
$\nu$-integration in the lower half-plane. This corresponds to the choice
$w_1<1$ and $w_2>1$, which is compatible with the symmetry $w_1\leftrightarrow
  \frac{1}{w_2}$. In \cite{Bartels:2011ge}, this calculation was carried out
  for the integrals appearing at two loops, with the result
\begin{equation}
	\mathcal{I}_7\left[E_{\nu,n}\right]=\frac{1}{2}\left(\log\left|1+\frac{w_1}{1+\frac{1}{w_2}}\right|^2+\log\left|1+\frac{1+\frac{1}{w_2}}{w_1}\right|^2\right),
	\label{eq:7pt_2l}
\end{equation}
which takes the form of the six-point two-loop result with a rescaled variable,
$w\rightarrow \frac{w_1}{1+\frac{1}{w_2}}$.  From this result, we can also
immediately obtain $\mathcal{I}_7\left[E_{\mu,m}\right]$ by making use of the
symmetry \eqref{eq:sym_munu1} discussed before.
Indeed, a special class of integrals is obtained when only one of the two
energy eigenvalues $E_{\nu,n}$ or $E_{\mu,m}$ appears in the integrand, for
definiteness let us choose $E_{\nu,n}$. In this case, the $\mu$-integration can
be carried out explicitly and results in a rescaling of the parameters
$(w_1,w_1^\ast)$. This reduces the integral to a six-point
integral, that is, upon replacing 
\begin{equation}
  w \to \frac{w_1}{1+\frac{1}{w_2}} 
\end{equation}
one can effectively obtain the simple two-loop solution \eqn{eq:7pt_2l} from
the corresponding two-loop LLA integral of \eqn{eqn:master6}.  Starting from
three loops, integrals with both types of energy eigenvalues $E_{\nu,n}$ and
$E_{\mu,m}$ appear causing a more complicated result.  We therefore have to
resort to other means of solving the integral. As in the six-point case, a
sensible starting point is trying to understand the relevant functions
describing the remainder function.

\subsection{Harmonic polylogarithms in two variables}
\label{ssec:2dHPL}

Contrary to the situation in the six-point scenario, where the kinematics is
determined by one complex parameter $w$, in the seven-point case we need to 
find SVHPLs in two complex variables $w_1$ and $w_2$ (cf.~\eqn{eq:7pt_2l}).
In the discussion to follow we will use variables $y$ and $z$, which will
be related to $w_1$ and $w_2$ later on.

\subsubsection{Two new letters}

Harmonic Polylogarithms depending on two complex parameters -- or
two-dimensional harmonic polylogarithms (2dHPLs) for short\footnote{In the
following sections we will sometimes refer to the HPLs and SVHPLs discussed in
\subsecref{ssec:1dsvhpl} as 1dHPLs and 1dSVHPLs, respectively.} -- have been
constructed in \rcite{Gehrmann:2000zt}. The implementation relies on
introducing two new integration weights accompanying the weights $f_a$ defined
in \eqn{eqn:intweight},
\begin{equation}
  \label{eqn:intweight2}
  f_{z}=\frac{1}{t+z}\quad\textrm{and}\quad 
  f_{1-z}=\frac{1}{1-t-z} \,.
\end{equation}
Similar to the six-point scenario, where the function $f_{-1}$ does not appear,
the function $f_{z}$ is not needed for the seven-point remainder function in
MRK.  In accordance, we will introduce the additional letter $x_{1-z}$ only. 

Up to weight three, all 2dHPLs can be expressed in terms of generalized
polylogarithms. For the simplest cases the relations read:
\begin{align}
  H_{1-z}(y) &= H_1\Big(\frac{y}{1-z}\Big)=-\log\Big(1-\frac{y}{1-z}\Big),\nnl
  H_{1,1-z}(y) &= \frac{1}{2}\log^2(1-y)-\log(1-y)\log(1-z)+\Li_2\Big(\frac{z}{1-y}\Big)-\Li_2(z)\,,
\end{align}
where a complete set of relations leading to expressions for all possible
labels up to weight three is presented in appendix A.2 of
\rcite{Gehrmann:2000zt}. 

While the labels $y$ and $z$ appear to be on unequal footing in the above
formul\ae, this is actually a choice of notation only: there are numerous
relations between different representations of 2dHPLs. In particular, it is
always possible to switch $y$ and $z$ in label and argument of a 2dHPL, for
example:
\begin{equation}
  \label{eqn:switchrel}
  H_{0, 1, \oy}(z) = 
  H_1(z) H_{0,1}(y) - H_1(y) H_{0,1}(z) - 
H_{0, 1, 1}(y) + H_{0, 1, 1}(z) + H_{0, 1, 1 - z}(y)\,.
\end{equation}
This type of relation, which can be easily derived for every label by reverting
to the integral representation of 2dHPls, will be a crucial ingredient in
fixing $\zm$-terms in 2dSVHPLs below. 

In order to have a canonical representation, we will choose 2dHPLs with labels
from $\{0,1,1-z\}$ for the argument $y$ and 1dHPLs with labels from $\{0,1\}$
for the argument $z$.  Solving for shuffle relations by choosing a Lyndon basis
\cite{Blumlein:2003gb} for the labels we will finally use 
\begin{equation}
  \label{eqn:2dHPLbasis}
  H_{\mathrm{Lyndon}(\lbrace 0,1,\oz\rbrace)}(y)\quad\textrm{and}\quad H_{\mathrm{Lyndon}(\lbrace 0,1\rbrace)}(z)\,.
\end{equation}
As pointed out in \rcite{Bargheer:2015djt}, this choice is actually a
basis for the 2dHPLs appearing in the seven-point remainder function in MRK.


\subsubsection{Differential equations for 2dHPLs}

With a second complex parameter entering the definition of 2dHPLs, an obvious
question is the one about the differential behavior of those functions. While
one could use relations like \eqn{eqn:switchrel} and thus trace back
derivatives with respect to one variable appearing in the label of the polylogarithm
to a derivative with respect to the argument, it is far more efficient to
consider derivatives with respect to the labels of a 2dHPL directly. The
necessary formul\ae{} for taking those derivatives are listed and explained in
\appref{app:derivatives}.

In terms of a generating function of 2dHPLs\footnote{Again, instead of noting
the full word $s$ in the subscript of a
2dHPL, we just write the indices of the letters.} with argument $y$ 
\begin{align}
  H^{\lbrace 0,1,\oz\rbrace}(y)=&\sum\limits_{s\in X(\lbrace x_0,x_1,x_{\oz}\rbrace)}H_s(y)s\nnl
  =&\,1+H_0(y)\,x_0+H_1(y)\,x_1+H_{\oz}(y)\,x_{\oz}\nnl
  &\quad+H_{0,0}(y)\,x_0x_0+H_{0,1}(y)\,x_0x_1+H_{0,\oz}(y)\,x_0x_{\oz}\nnl
  &\quad+H_{1,0}(y)\,x_1x_0+H_{1,1}(y)\,x_1x_1+H_{1,\oz}(y)\,x_1x_{\oz}\nnl
  &\quad+H_{\oz,0}(y)\,x_{\oz}x_0+H_{\oz,1}(y)\,x_{\oz}x_1+H_{\oz,\oz}(y)\,x_{\oz}x_{\oz}\nnl
  &\quad+\dots\,,
  \label{eqn:genfunc}
\end{align}
the $y$-derivative can be written down immediately after considering the
defining equation \ref{eqn:polylogdef} together with the additional integration
weight $f_{1-z}$:
\begin{equation}
  \label{eqn:yderivative}
  \frac{\pd}{\pd y}H^{\lbrace 0,1,\oz\rbrace}(y)=\Bigg(\frac{x_0}{y}+\frac{x_1}{1\!-\!y}+\frac{x_{\oz}}{1\!-\!y\!-\!z}\Bigg)H^{\lbrace 0,1,\oz\rbrace}(y)\,.
\end{equation}
Proceeding to the derivative with respect to $z$, it is no longer possible to
write the derivative in multiplicative form as in \eqn{eqn:yderivative}. Instead,
one can describe the pattern of how letters and prefactors are attached to
existing words depending on their particular letters. Writing  
\begin{equation}
  \label{eqn:zderivative}
  \frac{\pd}{\pd z}H^{\lbrace 0,1,\oz\rbrace}(y)=\Xi\Big(H^{\lbrace 0,1,\oz\rbrace}(y)\Big)\,
\end{equation}
the operation $\Xi$ acts as follows:
\begin{subequations}
  \label{eqn:Xi}
\begin{itemize}
\item for each sequence of letters $s=x_{\oz}\ldots x_{\oz}$, promote
  \begin{equation}
    \label{rule:1}
    s\quad\rightarrow\quad 
        \frac{x_0s}{\oz}+\frac{x_1s}{z}-\frac{sx_0}{\oz}-\frac{sx_1}{z}
  \end{equation}
\item for each sequence of letters $s=x_1\ldots x_1$, promote
  \begin{equation}
    \label{rule:2}
    s\quad\rightarrow\quad 
    \frac{x_{\oz}s}{z(1-z)}-\frac{sx_{\oz}}{z(1-z)}
  \end{equation}
\item to any complete word $s$, add a leading $\oz$:
  \begin{equation}
    \label{rule:3}
    s\quad\rightarrow\quad 
    \frac{y}{(1-z)(1-y-z)}\,x_{\oz}s\,.
  \end{equation}
\end{itemize}
\end{subequations}
In order to extract the derivative of a particular 2dHPL with label $s$ one
expands both sides of \eqn{eqn:zderivative} and compares the coefficients of
the word $s$. 

In practice, aiming to find the $z$-derivative of $H_{0,\oz,1}(y)$ for example,
one has to browse through all words of length two, which upon adding one letter
following the rules \eqn{eqn:Xi} will yield the word $x_0x_{\oz}x_1$. Starting
from $H_{0,1}(y)x_0x_1$, one can insert a letter $x_{\oz}$ between $x_0$ and
$x_1$, making use of the first part of rule (\ref{rule:2}). Taking
$H_{0,\oz}(y)x_0x_{\oz}$, a letter $x_1$ can be appended to the right, which
amounts to using the last part of rule (\ref{rule:1}). Finally, the word
$x_0x_{\oz}x_1$ can be reached by prepending $x_0$ to the word $x_{\oz}x_1$
accompanying $H_{\oz,1}(y)$, thus using the first term in rule (\ref{rule:1}):
\begin{align}
  \frac{\pd}{\pd z}H_{0,\oz,1}(y)x_0x_{\oz}x_1&=\frac{H_{0,1}(y)}{z(\oz)}x_0\tblue{x_{\oz}}x_1-\frac{H_{0,\oz}(y)}{z}x_0x_{\oz}\tblue{x_1}+\frac{H_{\oz,1}(y)}{\oz}\tblue{x_0}x_{\oz}x_1\,.
\end{align}
Another example, where one has to make use of rules (\ref{rule:1}) and
(\ref{rule:3}) reads:
\begin{align}
  \frac{\pd}{\pd z}H_{\oz,0,0}(y)x_{\oz}x_0x_0&=\frac{y}{(1\!-\!z)(1\!-\!y\!-\!z)}H_{0,0}(y)\,\tblue{x_{\oz}}x_0x_0-\frac{1}{\oz}H_{1-z,0}(y)x_{\oz}\tblue{x_0}x_0\,.
\end{align}
%


\subsection{Single-valued harmonic polylogarithms in two variables}
\label{ssec:2dsvHPL}

The canonical way to identify single-valued versions of 2dHPLs would be to find
a generalization of the single-valuedness condition formulated in
\rcite{Brown2004527} for the alphabet $\lbrace x_0,x_1 \rbrace$. However,
although this generalization does most certainly exist, an explicit expression
thereof is currently not known to us\footnote{As pointed out in the
conclusions, the paper \cite{DelDuca:2016lad} provides an explicit construction
of those functions.}. Therefore we continue on a different path: we postulate
several constraints the single-valued versions of 2dHPLs should satisfy and
later on argue that the functions thus constructed are indeed single-valued. In
order to find those constraints, we take guidance by the properties of 1dHPLs
reviewed in \subsecref{ssec:1dsvhpl}:
\begin{itemize}
  \item \textbf{Differential equations:} 1dSVHPLs satisfy the same differential
    equations as their 1dHPL counterpart (cf.~\eqns{eqn:diffhpl}{eqn:svhpldiff}): therefore we require
    the generating functional of 2dSVHPLs with argument $y$
    \begin{align}
      \CL^{\lbrace 0,1,\oz\rbrace}(y)=&\sum\limits_{s\in X(\lbrace x_0,x_1,x_{\oz}\rbrace)}\CL_s(y)s\nnl
      =&\,1+\CL_0(y)\,x_0+\CL_1(y)\,x_1+\CL_{\oz}(y)\,x_{\oz}\nnl
      &\quad+\CL_{0,0}(y)\,x_0x_0+\CL_{0,1}(y)\,x_0x_1+\CL_{0,\oz}(y)\,x_0x_{\oz}\nnl
      &\quad+\CL_{1,0}(y)\,x_1x_0+\CL_{1,1}(y)\,x_1x_1+\CL_{1,\oz}(y)\,x_1x_{\oz}\nnl
      &\quad+\CL_{\oz,0}(y)\,x_{\oz}x_0+\CL_{\oz,1}(y)\,x_{\oz}x_1+\CL_{\oz,\oz}(y)\,x_{\oz}x_{\oz}\nnl
      &\quad+\dots\,,
    	\label{eqn:genfuncCL}
    \end{align}
    to satisfy
    \begin{align}
	    \frac{\pd}{\pd y}\CL^{\lbrace 0,1,\oz\rbrace}(y)&=\Bigg(\frac{x_0}{y}+\frac{x_1}{1-y}+\frac{x_{\oz}}{1-y-z}\Bigg)\CL^{\lbrace 0,1,\oz\rbrace}(y)\label{eqn:2dsvhpldiffy}\\
      \frac{\pd}{\pd z}\CL^{\lbrace 0,1,\oz\rbrace}(y)&=\Xi\Big(\CL^{\lbrace 0,1,\oz\rbrace}(y)\Big)\,.\label{eqn:2dsvhpldiffz}
    \end{align}
    where the operation $\Xi$ has been defined in \eqn{eqn:Xi}.
  \item \textbf{Limiting behavior:} In the limit $z\to 0$, the alphabet will
    shrink since $x_{\oz}\to x_1$. Thus we demand to recover the corresponding
    1dSVHPLs in the limit $(1-z)\rightarrow 1$.
  \item \textbf{Switching variables:} Consistency with the relations switching
    the variable in the label and the argument. This means we require all
    relations like \eqn{eqn:switchrel} to hold upon replacing $H\rightarrow
    \CL$.
\end{itemize}
In short, we require 2dSVHPLs to inherit the properties of 2dHPLs
and in addition we demand a consistent reduction to 1dSVHPLs in the limit $z\to
0$. We will explicitly show in the following that those constraints are indeed
sufficient to pin down 2dSVHPLs in two variables to at least weight four.

In practice, we can gain some experience regarding the structure of the
2dSVHPLs by studying a simple ad hoc construction: We start from
the 2dHPLs as given by Gehrmann and Remiddi and reviewed in
\subsecref{ssec:2dHPL}.  Since these functions can be expressed in terms of
generalized polylogarithms up to weight three, we can promote each 1dHPL
separately to its single-valued version using relations like \eqn{eqn:svhpl}. 

This is most easily explained using an example: a candidate for a single-valued
2dHPL can be obtained via
\begin{align}
  H_{1,1-z}(y) &= 
   \frac{1}{2}\log^2(1-y)-\log(1-y)\log(1-z)+\Li_2\Big(\frac{z}{1-y}\Big)-\Li_2(z)\nnl
   &=H_{1,1}(y)-H_1(y)H_1(z)+H_{0,1}\Big(\frac{z}{1\!-\!y}\Big)-H_{0,1}(z)\nnl
   &\rightarrow \CL_{1,1}(y)-\CL_1(y)\CL_1(z)+\CL_{0,1}\Big(\frac{z}{1\!-\!y}\Big)-\CL_{0,1}(z).
   \label{eqn:example_2dsvhpl}
\end{align}
Making use of the scaling relation
\begin{align}
	H_{a_1,a_2, \ldots,a_n}\left(\frac{y}{1-z}\right)=H_{(1-z)a_1,(1-z)a_2, \ldots, (1-z)a_n}(y),
	\label{eqn:rescale_ident}
\end{align}
where $a_i\in \{0,1\}$ and which holds whenever the last index is not $0$ (see
the discussion around \eqn{eqn:scaling}), as well as
\begin{align}
	H_{\underbrace{\scriptstyle{0, \ldots, 0}}_n}\left(\frac{y}{1-z}\right)=\frac{1}{n!}\left(H_0(y)+H_1(z)\right)^n\,,
\end{align}
one can express the above candidate for a single-valued polylogarithm in terms
of 2dHPLs:
\begin{equation}
  \label{eqn:representation}
  \CL_{1,\oz}(y)=H_{1,\oz}(y)+H_1(y)H_{\ozb}(\yb)+H_{\ozb,1}(\yb)+\CL_0(z)\Big(H_{\ozb}(\yb)-H_1(\yb)\Big)-\CL_1(z)H_1(\yb)\,.
\end{equation}
Note that in \eqn{eqn:representation} the variable $z$ in the label is
complex-conjugated whenever the argument of the HPL is $\bar{y}$.  Following
this ad hoc approach we find functions up to weight three which perfectly match
the first orders of the integral \eqn{eq:master7}.  Furthermore, note that we
did not express 1dSVHPLs of $z$ in terms of usual HPLs, as this relates to a
feature of 2dSVHPLs to be elaborated on below: if expressed in the basis
\eqn{eqn:2dHPLbasis}, 2dSVHPLs split into a \textit{canonical} part as well as
a part in which 1dSVHPLs of argument $z$ are multiplied by 2dHPLs of arguments
$y$ and $\yb$ (with labels possibly containing $\oz$ and $\ozb$).  By
\textit{canonical} we refer to the pattern
\begin{equation}
  \left.\CL_{a_1,a_2, \ldots, a_n}(y)\right|_{\mathrm{can.}}=\sum\limits_{k=0}^n H_{a_1, \ldots, a_k}(y)H_{\bar{a}_n, \bar{a}_{n-1}, \ldots ,\bar{a}_{k+1}}(\bar{y})
	\label{eq:def_canonical}
\end{equation}
that is already present for 1dSVHPLs (see \eqn{eqn:svhpl}).
Quite naturally, the additional terms one finds beyond the canonical part are
exactly those needed to preserve the derivative rule \eqn{eqn:2dsvhpldiffz}.

Since our ad hoc construction of 2dSVHPLs only works up to weight three, we
would now like to turn these observations into a construction of higher-weight
2dSVHPLs by the following algorithm: We start from a known 2dSVHPL and add a
letter to the left by integrating in $y$, thus making use of the differential
equation (\ref{eqn:2dsvhpldiffy}). This fixes the 2dSVHPL of higher weight up
to a function of $\bar{y}$, $z$ and $\bar{z}$.

Based on the assumptions that HPLs of argument $z$ only appear in single-valued
combinations we make the most general ansatz of terms
\begin{equation}
	\CL_{a_1,\ldots,a_n}(z)H_{b_1,\ldots,b_m}(\bar{y}),
	\label{eqn:ansatz_extra_terms}
\end{equation}
where $a_i\in\{0,1\}$ and $b_i\in\{0,1,\ozb\}$,
compatible with the overall weight.
Demanding that
this ansatz satisfies relations \eqn{eqn:2dsvhpldiffz} for differentiation in
$z$ then fixes the ansatz completely. 

To clarify the procedure, let us consider an example.
Starting from the obvious weight one expressions
\begin{align}
	\CL_0(y)&=H_0(y)+H_0(\bar{y}),\nnl
	\CL_1(y)&=H_1(y)+H_1(\bar{y}),\nnl
	\CL_{\oz}(y)&=H_{1-z}(y)+H_{\ozb}(\bar{y}),
\end{align}
we can, for example, write down an ansatz for $\CL_{1,1-z}(y)$ as
\begin{align}
	\CL_{1,1-z}(y)=\,&H_{1,1-z}(y)+H_{\ozb, 1}(\yb)+H_1(y)H_{\ozb}(\yb)\nnl
	& +\CL_0(z)\left(c_1 H_0(\yb)+c_2H_1(\yb)+c_3H_{\ozb}(\yb)\right)\nnl
	& +\CL_1(z)\left(c_4 H_0(\yb)+c_5H_1(\yb)+c_6H_{\ozb}(\yb)\right),
	\label{eq:ansatz_11mz}
\end{align}
which by differentiation with respect to $z$ is then fixed to give
\eqn{eqn:representation}.  Carrying this out for all functions up to weight
three, we can compare the results of this algorithm with our ad hoc
construction and find a perfect matching, as we should.  Additional
complications start from weight four where $\zeta$-values are going to appear.
As the 2dSVHPLs at weight three do not contain any zetas, those terms cannot be
fixed by the $y$- and $z$-derivative and we have to add a term
\begin{equation}
	\zeta_3\Big(c_1 H_1(\yb)+c_2 H_{\ozb}(\yb)\Big)
	\label{eq:zeta_ansatz}
\end{equation}
to the ansatz of every 2dSVHPLs at weight four. Note that the two terms above
are the only ones consistent with the reduction $z\rightarrow 0$, as well as
vanishing in the limit $\yb\rightarrow 0$.  Demanding a consistent reduction in
the limit $z\rightarrow 0$ and consistency with the relations exchanging
argument and label fixes most of the $\zeta$-terms, but not all.  However, we
can impose an additional constraint: as shown in \cite{Gehrmann:2000zt}, a
2dHPL evaluated at $y=1-z$ can be written as a combination of 1dHPLs of
argument $z$.  Similarly, setting $y=1-z$, $\yb=\ozb$ in our ansatz, we expect
to obtain a combination of 1dSVHPLs of weight four. As it turns out, this
constraint is strong enough to fix all coefficients.  As an example containing
a $\zeta$-value, we find
\begin{align}
	\CL_{0,0,1,1-z}(y)=&\left.\CL_{0,0,1,1-z}(y)\right|_{\mathrm{can.}}+\CL_{0,0,1}(z)H_{\ozb}(\yb)+\CL_{0,1}(z)\Big(H_0(y)H_{\ozb}(\yb)+H_{\ozb,0}(\yb)\Big)  \nnl
	&+\CL_{1}(z)\Big(-H_{0,0}(y)H_1(\yb)-H_0(y)H_{1,0}(\yb)-H_{1,0,0}(\yb)\Big)\nnl
	&+\CL_{0}(z)\Big(-H_{0,0}(y)H_1(\yb)+H_{0,0}(y)H_{\ozb}(\yb)-H_0(y)H_{1,0}(\yb)+H_0(y)H_{\ozb,0}(\yb)\nnl
	&\qquad\qquad\quad-H_{1,0,0}(\yb)+H_{\ozb,0,0}(\yb)\Big)-2\zeta_3H_{\ozb}(\yb).
\end{align}
The expressions for all other Lyndon basis elements up to weight four can be
found in the file attached to the arXiv submission of this paper.

Going on to weight five, our constraints do not seem to be strong enough to fix
all coefficients, which is due to the growth of both the number of Lyndon basis
elements and the larger number of terms appearing in the ansatz for the
$\zeta$-terms, i.e.~the analogue of \eqn{eq:zeta_ansatz} at weight five.  We
are only able to fully fix those 2dSVHPLs at weight five whose label contains
two different indices only\footnote{This is not surprising, as those 2dSVHPLs
can be constructed from 1dSVHPLs using the rescaling identity.}.  The
expressions for the weight-five 2dSVHPLs can also be found in the attached
file, but note that those still contain fudge factors. It would be interesting
to see if there are additional constraints which allow to completely fix the
functions at weight five as well.

Up to weight three it is obvious from our ad hoc construction that the
resulting functions are single-valued: they are composed from single-valued
components by definition. For higher weights we can only argue that this is
indeed the case: Starting from our ansatz and fixing all fudge coefficients
does not only reproduce the 2dSVHPLs constructed na\"{i}vely, but yields
functions, which including their $\zm$-parts perfectly match the analytical
properties of the integral \eqn{eq:master7}. While this does not prove
single-valuedness, the perfect matching with the explicit calculation of the
integral strongly supports our conjecture.


\subsection{Matching the results}
\label{ssec:results}

Now that we have constructed a suitable class of functions, we want to generate
expressions for the seven-point remainder function \eqn{eq:master7} beyond two
loops.  We do this by simply writing down an ansatz with the correct weight and
matching the series expansion of the ansatz to data generated from calculating
residues of \eqn{eq:master7}. This also allows us to identify the variables
in argument and label and we find that
\begin{equation}
	(y,z)\rightarrow \left(-w_1,-\frac{1}{w_2}\right)
\end{equation}
is the correct choice. In the following we will use the abbreviation $x:=\frac{1}{w_2}$.

This leads to the following results at two loops,
\begin{align}
	\mathcal{I}_7 \left[E_{\nu,n} \right]=&\,\frac{1}{2} \CL_{1}(-x)\CL_{1+x}(-w_1)+\frac{1}{2}\CL_{0,1+x}(-w_1)+\frac{1}{2}\CL_{1+x,0}(-w_1)+\CL_{1+x,1+x}(-w_1),\nnl 
	\mathcal{I}_7 \left[E_{\mu,m}\right]=&-\frac{1}{2} \CL_{1}(-w_1) \CL_{0}(-x)-\frac{1}{2} \CL_{1}(-w_1)\CL_1(-x)+\frac{1}{2} \CL_0(-x) \CL_{1+x}(-w_1)\nnl
	&+\CL_{1}(-x)\CL_{1+x}(-w_1)-\frac{1}{2} \CL_{1,1+x}(-w_1)-\frac{1}{2}\CL_{1+x,1}(-w_1)+\CL_{1+x,1+x}(-w_1)\nnl
	&+\frac{1}{2}\CL_{0,1}(-x)+\frac{1}{2}\CL_{1,0}(-x)+\CL_{1,1}(-x).\nonumber
\end{align}
Note again that $\mathcal{I}_7\left[E_{\mu,m}\right]$ can be obtained from
$\mathcal{I}_7\left[E_{\nu,n}\right]$ by using the symmetry
\eqref{eq:sym_munu1} as well as the relations \eqref{eqn:switchrel}.
At three loops we find
\begin{align}
	\mathcal{I}_7 \left[E_{\nu,n}^2\right]=&\,\frac{1}{2} \CL_{1}(-x) \CL_{0,1+x}(-w_1)+\frac{1}{4} \CL_{1,1}(-x)\CL_{1+x}(-w_1)+\frac{1}{4} \CL_{1}(-x)\CL_{1+x,0}(-w_1)\nnl
	&+\CL_{1}(-x) \CL_{1+x,1+x}(-w_1)+\frac{1}{4}\CL_{0,0,1+x}(-w_1)+\frac{1}{2}\CL_{0,1+x,0}(-w_1)+\CL_{0,1+x,1+x}(-w_1)\nnl
	&+\frac{1}{4}\CL_{1+x,0,0}(-w_1)+\CL_{1+x,0,1+x}(-w_1)+\CL_{1+x,1+x,0}(-w_1)+2 \CL_{1+x,1+x,1+x}(-w_1)\nonumber
\end{align}
as well as
\begin{align}
	\mathcal{I}_7\left[E_{\nu,n}\,E_{\mu,m}\right]=&-\frac{1}{4} \CL_{0,1}(-w_1) \CL_{0}(-x)-\frac{1}{4}\CL_{0,1}(-w_1) \CL_{1}(-x)-\frac{3}{4} \CL_{0,1}(-w_1)\CL_{1+x}(-w_1)\nnl
	&-\frac{1}{4} \CL_{1}(-w_1)\CL_{0,1}(-x)+\frac{1}{2} \CL_{0,1}(-x) \CL_{1+x}(-w_1)+\frac{1}{4}\CL_{0}(-x) \CL_{0,1+x}(-w_1)\nnl
	&+\frac{1}{2} \CL_{1}(-x)\CL_{0,1+x}(-w_1)-\frac{1}{4} \CL_{1,0}(-w_1)\CL_{0}(-x)-\frac{1}{4} \CL_{1,0}(-w_1) \CL_{1}(-x)\nnl
	&-\frac{3}{4}\CL_{1,0}(-w_1) \CL_{1+x}(-w_1)+\frac{1}{4} \CL_{1,0}(-x)\CL_{1+x}(-w_1)-\frac{1}{2} \CL_{1,1}(-w_1)\CL_{0}(-x)\nnl
	&-\frac{1}{2} \CL_{1,1}(-w_1) \CL_{1}(-x)-\frac{1}{4}\CL_{1}(-w_1) \CL_{1,1}(-x)+\CL_{1,1}(-x)\CL_{1+x}(-w_1)\nnl
	&-\frac{1}{4} \CL_{1}(-x)\CL_{1,1+x}(-w_1)+\frac{1}{4} \CL_{0}(-x)\CL_{1+x,0}(-w_1)+\frac{1}{2} \CL_{1}(-x)\CL_{1+x,0}(-w_1)\nnl
	&+\frac{1}{2} \CL_{0}(-x)\CL_{1+x,1+x}(-w_1)+\frac{3}{2} \CL_{1}(-x)\CL_{1+x,1+x}(-w_1)+\frac{1}{2} \CL_{0,1,1+x}(-w_1)\nnl
	&+\frac{1}{2}\CL_{0,1+x,1}(-w_1)+\frac{1}{2} \CL_{0,1+x,1+x}(-w_1)+\frac{1}{4}\CL_{1,0,1+x}(-w_1)-\frac{1}{2} \CL_{1,1,1+x}(-w_1)\nnl
	&+\frac{1}{2}\CL_{1,1+x,0}(-w_1)-\frac{1}{2} \CL_{1,1+x,1+x}(-w_1)+\frac{1}{4}\CL_{1+x,0,1}(-w_1)+\CL_{1+x,0,1+x}(-w_1)\nnl
	&+\frac{1}{2}\CL_{1+x,1,0}(-w_1)-\frac{1}{2} \CL_{1+x,1,1}(-w_1)+\frac{1}{2}\CL_{1+x,1+x,0}(-w_1)-\frac{1}{2} \CL_{1+x,1+x,1}(-w_1)\nnl
	&+2\CL_{1+x,1+x,1+x}(-w_1).\nonumber
\end{align}
As explained before, the remaining integral at three loops, $\mathcal{I}_7\left[E_{\mu,m}^2\right]$, can be obtained by symmetry.
All further results up to five loops are too lengthy to be reproduced here and
we refer the reader to the file accompanying this paper.

Before we close let us make one additional but important remark.  Recall that
the key property of the one-dimensional SVHPLs that allowed us to directly
construct the full result from a small set of residues in the six-point was
that the leading term of the 1dSVHPL was a HPL with the same index structure
which only depends on $w$.  However, comparing with the explicit expressions of
our 2dSVHPLs, e.g.~\eqn{eqn:representation}, we see that a similar statement
holds -- the leading term of a given 2dSVHPL is the 2dHPL with the same index
structure which only depends on $y$ and $z$ but not on the complex-conjugated
variables.  This means that we should be able to construct the full result
solely from the residues at
\begin{equation}
	\nu=-\frac{i\,n}{2},\quad\mathrm{and}\quad\mu=\frac{i\,m}{2},
	\label{eq:necessary_residues}
\end{equation}
which leads to an expression in 2dHPLs, and then making the replacement
\begin{equation}
	H_{w}(y)\rightarrow \CL_{w}(y).
	\label{eq:lift_2dhpls}
\end{equation}
We have checked that this indeed reproduces the full result.  This in turn
means that it should be possible to follow a procedure similar to
\cite{Broedel:2015nfp} to reconstruct the remainder function from a set of
simple basis integrals. 

It is because of this remark that we present the formula for the five-loop
remainder function in terms of 2dSVHPLs in the attached file, even though we
cannot fully fix the $\zeta$-parts of all those functions yet.  We have
obtained the full result in 2dHPLs and checked that the prescription
\eqn{eq:lift_2dhpls} works for the $\zeta$-free part. Furthermore, the
integrals contributing to the five-loop remainder function which only contain
one kind of energy eigenvalue,
\begin{equation}
	\mathcal{I}_7\left[E_{\nu,n}^4\right],\,\mathcal{I}_7\left[E_{\mu,m}^4\right]
\end{equation}
only contain 2dSVHPLs of weight five with two different indices, which we fully
understand.  Here, too, the prescription \eqn{eq:lift_2dhpls} works.  We are
therefore convinced that the formula as written down in the attached file is
correct.


\section{Conclusions}
\label{sec:conclusion}

Setting up an efficient approach for calculating the MHV remainder function for
seven points in $\CN=4$ super-Yang--Mills requires the construction of
single-valued harmonic polylogarithms in two variables, 2dSVHPLs. In this paper, 
we have started the investigation of their analytical properties and constructed
those functions up to and including weight four. The analytical constraints we 
are using, however, are not strong enough for completely determining the
single-valued version of 2dHPLs starting from weight five, since we cannot fix
the coefficients of all terms proportional to zeta values.

Upon availability of expressions for 2dHPLs, it is possible to apply a similar
concept as the one introduced in \rcite{Drummond:2015jea}: by calculating a
certain subset of the residues contributing to the seven-point MHV remainder
function only, one can determine the leading term of the 2dSVHPLs, which later
on can be promoted to the full single-valued expression.  
%
%

Using this method, we have expressed the remainder function in terms of
2dSVHPLs up to five loops.
It would be interesting to see whether there are further constraints
on the 2dSVHPLs which would allow us to go beyond weight five.

While we provide an ad-hoc construction of the 2dSVHPLs, in a recent paper
\cite{DelDuca:2016lad} an explicit construction of those functions is provided
to arbitrary weight in a different language. Naturally, it would be interesting
to compare our results to the more general approach in ref.~\cite{DelDuca:2016lad}.

Furthermore, the pattern we find is simple enough to suggest that it should be
possible to identify a formula similar to the formula derived in
\rcite{Pennington:2012zj} for the LLA of the six-point remainder function.  

\subsection*{Acknowledgments} 
We would like to thank Claude Duhr, Falko Dulat, Matteo Rosso and Till Bargheer
for helpful discussions.  The research of JB is supported in part by the SFB
647 “Raum--Zeit--Materie. Analytische und Geometrische Strukturen” and the Marie
Curie Network GATIS (gatis.desy.eu) of the European Union’s Seventh Framework
Programme FP7-2007-2013 under grant agreement No.~317089.  The work of MS is
partially supported by the Swiss National Science Foundation through the NCCR
SwissMap.


\section*{Appendix}
\appendix

\section{Derivatives of harmonic polylogarithms}
\label{app:derivatives}

Derivatives of HPLs can be most easily stated in general form using the
language of Goncharov polylogarithms. Given their integral definition
\begin{equation}
  G_{a_1,a_2,\ldots,a_n}(y)=\int_0^y \frac{\dd t}{t-a_1}G_{a_2,\ldots,a_n}(t)\,,
\end{equation}
and comparing with the integration weights $f_0$, $f_1$ and $f_{1-z}$ leads to 
\begin{equation}
  G_{a_1,a_2,\ldots,a_n}(y)=(-1)^{(\#(1)+\#(1-z))} H_{a_1,a_2,\ldots,a_n}(y)\,,
\end{equation}
that is, the relative sign is determined by the number of $1$'s and $(1-z)$'s
appearing in the label. In terms of Goncharov polylogarithms, derivatives with
respect to label and argument read \cite{Goncharov:2001iea}:
\begin{align}
	\frac{\pd}{\pd y}G_{a_1,a_2,\ldots,a_n}(y)&=\frac{1}{y-a_1}G_{a_2,\ldots,a_n}(y)\nnl
	\frac{\pd}{\pd a_i}G_{a_1,a_2\ldots,a_n}(y)&=\frac{1}{a_{i-1}-a_i}G_{a_1,\ldots,\hat{a}_{i-1},\ldots,a_n}(y) \ + \ \frac{1}{a_i-a_{i+1}}G_{a_1,\ldots,\hat{a}_{i+1},\ldots,a_n}(y)\nnl
	&-\frac{a_{i-1}-a_{i+1}}{(a_{i-1}-a_i)(a_i-a_{i+1})}G_{a_1,\ldots,\hat{a}_i,\ldots,a_n}(y)
  \nnl
  \nnl
  \frac{\pd}{\pd a_n}G_{a_1,\ldots,a_n}(y)&=\frac{1}{a_{n-1}-a_n}G_{a_1,\ldots\hat{a}_{n-1},a_n}(y)
  -\frac{a_{n-1}}{(a_{n-1}-a_n)a_n}G_{a_1,\ldots,a_{n-1}}(y)\,,
   \label{eqn:derivativeG}
\end{align}

where $\hat{a}$ denotes omission of the respective entry. Given the terms of
the form $\frac{1}{a_{i-1}-a_i}$ in the derivatives, it is obvious that for
neighboring elements of the same kind one will get divergent terms. However,
by shifting the entries in the label by a small value $\epsilon$ one can
safely determine the derivative and successively take $\epsilon\rightarrow 0$.
For example one finds:
\begin{align}
	\frac{\pd}{\pd z}H_{\oz,\oz,\oz}(y)&=-\frac{\pd}{\pd z}G_{\oz+\epsilon,\oz,\oz+\epsilon}(y)\Big|_{\epsilon\rightarrow 0}\nnl
	&=-\frac{y\,G_{\oz,\oz}(y)}{(1-z)(1-y-z)}\nnl
	&=-\frac{y\,H_{\oz,\oz}(y)}{(1-z)(1-y-z)}\,.
\end{align}
%


\begin{bibtex}[\jobname]

@article{DelDuca:2016lad,
      author         = "Del Duca, Vittorio and Druc, Stefan and Drummond, James
                        and Duhr, Claude and Dulat, Falko and Marzucca, Robin and
                        Papathanasiou, Georgios and Verbeek, Bram",
      title          = "{Multi-Regge kinematics and the moduli space of Riemann
                        spheres with marked points}",
      journal        = "JHEP",
      volume         = "08",
      year           = "2016",
      pages          = "152",
      doi            = "10.1007/JHEP08(2016)152",
      eprint         = "1606.08807",
      archivePrefix  = "arXiv",
      primaryClass   = "hep-th",
      SLACcitation   = "
}

@article{Drummond:2015jea,
      author         = "Drummond, J. M. and Papathanasiou, G.",
      title          = "{Hexagon OPE Resummation and Multi-Regge Kinematics}",
      journal        = "JHEP",
      volume         = "02",
      year           = "2016",
      pages          = "185",
      doi            = "10.1007/JHEP02(2016)185",
      eprint         = "1507.08982",
      archivePrefix  = "arXiv",
      primaryClass   = "hep-th",
      reportNumber   = "CERN-PH-TH-2015-183",
      SLACcitation   = "
}

@article{Blumlein:2003gb,
      author         = "Blumlein, Johannes",
      title          = "{Algebraic relations between harmonic sums and associated
                        quantities}",
      journal        = "Comput. Phys. Commun.",
      volume         = "159",
      year           = "2004",
      pages          = "19-54",
      doi            = "10.1016/j.cpc.2003.12.004",
      eprint         = "hep-ph/0311046",
      archivePrefix  = "arXiv",
      primaryClass   = "hep-ph",
      reportNumber   = "DESY-03-134, SFB-CPP-03-51",
      SLACcitation   = "
}

@article{Bargheer:2016eyp,
      author         = "Bargheer, Till",
      title          = "{Systematics of the Multi-Regge Three-Loop Symbol}",
      year           = "2016",
      eprint         = "1606.07640",
      archivePrefix  = "arXiv",
      primaryClass   = "hep-th",
      reportNumber   = "DESY-16-115",
      SLACcitation   = "
}

@article{Goncharov:2001iea,
      author         = "Goncharov, A. B.",
      title          = "{Multiple polylogarithms and mixed Tate motives}",
      year           = "2001",
      eprint         = "math/0103059",
      archivePrefix  = "arXiv",
      primaryClass   = "math.AG",
      SLACcitation   = "
}

@article{Bargheer:2015djt,
      author         = "Bargheer, Till and Papathanasiou, Georgios and Schomerus,
                        Volker",
      title          = "{The Two-Loop Symbol of all Multi-Regge Regions}",
      journal        = "JHEP",
      volume         = "05",
      year           = "2016",
      pages          = "012",
      doi            = "10.1007/JHEP05(2016)012",
      eprint         = "1512.07620",
      archivePrefix  = "arXiv",
      primaryClass   = "hep-th",
      reportNumber   = "DESY-15-257",
      SLACcitation   = "
}
@article{Dixon:2013eka,
      author         = "Dixon, Lance J. and Drummond, James M. and von Hippel,
                        Matt and Pennington, Jeffrey",
      title          = "{Hexagon functions and the three-loop remainder
                        function}",
      journal        = "JHEP",
      volume         = "12",
      year           = "2013",
      pages          = "049",
      doi            = "10.1007/JHEP12(2013)049",
      eprint         = "1308.2276",
      archivePrefix  = "arXiv",
      primaryClass   = "hep-th",
      reportNumber   = "CERN-PH-TH-2013-142",
      SLACcitation   = "
}

@article{Golden:2014xqf,
      author         = "Golden, John and Spradlin, Marcus",
      title          = "{An analytic result for the two-loop seven-point MHV
                        amplitude in $ \mathcal{N} $ = 4 SYM}",
      journal        = "JHEP",
      volume         = "08",
      year           = "2014",
      pages          = "154",
      doi            = "10.1007/JHEP08(2014)154",
      eprint         = "1406.2055",
      archivePrefix  = "arXiv",
      primaryClass   = "hep-th",
      reportNumber   = "BROWN-HET-1656",
      SLACcitation   = "
}

@article{CaronHuot:2011ky,
      author         = "Caron-Huot, S.",
      title          = "{Superconformal symmetry and two-loop amplitudes in
                        planar N=4 super Yang-Mills}",
      journal        = "JHEP",
      volume         = "12",
      year           = "2011",
      pages          = "066",
      doi            = "10.1007/JHEP12(2011)066",
      eprint         = "1105.5606",
      archivePrefix  = "arXiv",
      primaryClass   = "hep-th",
      SLACcitation   = "
}

@article{Drummond:2014ffa,
      author         = "Drummond, James M. and Papathanasiou, Georgios and
                        Spradlin, Marcus",
      title          = "{A Symbol of Uniqueness: The Cluster Bootstrap for the
                        3-Loop MHV Heptagon}",
      journal        = "JHEP",
      volume         = "03",
      year           = "2015",
      pages          = "072",
      doi            = "10.1007/JHEP03(2015)072",
      eprint         = "1412.3763",
      archivePrefix  = "arXiv",
      primaryClass   = "hep-th",
      reportNumber   = "CERN-PH-TH-2014-256, LAPTH-232-14",
      SLACcitation   = "
}

@article{Bartels:2014ppa,
      author         = "Bartels, J. and Schomerus, V. and Sprenger, M.",
      title          = "{Heptagon Amplitude in the Multi-Regge Regime}",
      journal        = "JHEP",
      volume         = "10",
      year           = "2014",
      pages          = "67",
      doi            = "10.1007/JHEP10(2014)067",
      eprint         = "1405.3658",
      archivePrefix  = "arXiv",
      primaryClass   = "hep-th",
      reportNumber   = "DESY-14-076",
      SLACcitation   = "
}

@article{Bartels:2014mka,
      author         = "Bartels, J. and Schomerus, V. and Sprenger, M.",
      title          = "{The Bethe roots of Regge cuts in strongly coupled $
                        \mathcal{N}=4 $ SYM theory}",
      journal        = "JHEP",
      volume         = "07",
      year           = "2015",
      pages          = "098",
      doi            = "10.1007/JHEP07(2015)098",
      eprint         = "1411.2594",
      archivePrefix  = "arXiv",
      primaryClass   = "hep-th",
      reportNumber   = "DESY-14-208",
      SLACcitation   = "
}

@article{Bartels:2014jya,
      author         = "Bartels, Jochen and Kormilitzin, Andrey and Lipatov, Lev
                        N.",
      title          = "{Analytic structure of the $n=7$ scattering amplitude in
                        $\mathcal{N}=4$ theory in multi-Regge kinematics:
                        Conformal Regge cut contribution}",
      journal        = "Phys. Rev.",
      volume         = "D91",
      year           = "2015",
      number         = "4",
      pages          = "045005",
      doi            = "10.1103/PhysRevD.91.045005",
      eprint         = "1411.2294",
      archivePrefix  = "arXiv",
      primaryClass   = "hep-th",
      reportNumber   = "DESY-14-171",
      SLACcitation   = "
}

@article{Bartels:2013jna,
      author         = "Bartels, Jochen and Kormilitzin, Andrey and Lipatov, Lev",
      title          = "{Analytic structure of the $n=7$ scattering amplitude in
                        $\mathcal{N}=4$ SYM theory in the multi-Regge kinematics:
                        Conformal Regge pole contribution}",
      journal        = "Phys. Rev.",
      volume         = "D89",
      year           = "2014",
      number         = "6",
      pages          = "065002",
      doi            = "10.1103/PhysRevD.89.065002",
      eprint         = "1311.2061",
      archivePrefix  = "arXiv",
      primaryClass   = "hep-th",
      reportNumber   = "DESY-13-209",
      SLACcitation   = "
}

@article{Broedel:2015nfp,
      author         = "Broedel, Johannes and Sprenger, Martin",
      title          = "{Six-point remainder function in multi-Regge-kinematics:
                        an efficient approach in momentum space}",
      journal        = "JHEP",
      volume         = "05",
      year           = "2016",
      pages          = "055",
      doi            = "10.1007/JHEP05(2016)055",
      eprint         = "1512.04963",
      archivePrefix  = "arXiv",
      primaryClass   = "hep-th",
      reportNumber   = "HU-EP-15-61, HU-MATHEMATIK-2015-15",
      SLACcitation   = "
}

@article{Anastasiou:2016hlm,
      author         = "Anastasiou, Charalampos and Duhr, Claude and Dulat, Falko
                        and Furlan, Elisabetta and Gehrmann, Thomas and Herzog,
                        Franz and Lazopoulos, Achilleas and Mistlberger, Bernhard",
      title          = "{CP-even scalar boson production via gluon fusion at the
                        LHC}",
      year           = "2016",
      eprint         = "1605.05761",
      archivePrefix  = "arXiv",
      primaryClass   = "hep-ph",
      SLACcitation   = "
}

@article{Gehr,
      author         = "Gehrmann, T. and Remiddi, E.",
}

@article{Gehrmann:2000zt,
      author         = "Gehrmann, T. and Remiddi, E.",
      title          = "{Two loop master integrals for $\gamma^\ast\rightarrow 3$ jets: The
                        Planar topologies}",
      journal        = "Nucl. Phys.",
      volume         = "B601",
      year           = "2001",
      pages          = "248-286",
      doi            = "10.1016/S0550-3213(01)00057-8",
      eprint         = "hep-ph/0008287",
      archivePrefix  = "arXiv",
      primaryClass   = "hep-ph",
      reportNumber   = "KA-TTP-00-20",
      SLACcitation   = "
}

@article{Dixon:2015iva,
      author         = "Dixon, Lance J. and von Hippel, Matt and McLeod, Andrew
                        J.",
      title          = "{The four-loop six-gluon NMHV ratio function}",
      journal        = "JHEP",
      volume         = "01",
      year           = "2016",
      pages          = "053",
      doi            = "10.1007/JHEP01(2016)053",
      eprint         = "1509.08127",
      archivePrefix  = "arXiv",
      primaryClass   = "hep-th",
      reportNumber   = "SLAC-PUB-16352, CALT-2015-049",
      SLACcitation   = "
}

@article{Hatsuda:2014oza,
      author         = "Hatsuda, Yasuyuki",
      title          = "{Wilson loop OPE, analytic continuation and multi-Regge
                        limit}",
      journal        = "JHEP",
      volume         = "10",
      year           = "2014",
      pages          = "38",
      doi            = "10.1007/JHEP10(2014)038",
      eprint         = "1404.6506",
      archivePrefix  = "arXiv",
      primaryClass   = "hep-th",
      reportNumber   = "DESY-14-057",
      SLACcitation   = "
}
	
@article{Lipatov:2012gk,
      author         = "Lipatov, Lev and Prygarin, Alexander and Schnitzer,
                        Howard J.",
      title          = "{The Multi-Regge limit of NMHV Amplitudes in N=4 SYM
                        Theory}",
      journal        = "JHEP",
      volume         = "01",
      year           = "2013",
      pages          = "068",
      doi            = "10.1007/JHEP01(2013)068",
      eprint         = "1205.0186",
      archivePrefix  = "arXiv",
      primaryClass   = "hep-th",
      reportNumber   = "BRX-TH-655, BROWN-HET-1634",
      SLACcitation   = "
}

@article{Dixon:2014iba,
      author         = "Dixon, Lance J. and von Hippel, Matt",
      title          = "{Bootstrapping an NMHV amplitude through three loops}",
      journal        = "JHEP",
      volume         = "10",
      year           = "2014",
      pages          = "065",
      doi            = "10.1007/JHEP10(2014)065",
      eprint         = "1408.1505",
      archivePrefix  = "arXiv",
      primaryClass   = "hep-th",
      reportNumber   = "SLAC-PUB-15970",
      SLACcitation   = "
}

@article{Moch:2001zr,
      author         = "Moch, Sven and Uwer, Peter and Weinzierl, Stefan",
      title          = "{Nested sums, expansion of transcendental functions and
                        multiscale multiloop integrals}",
      journal        = "J. Math. Phys.",
      volume         = "43",
      year           = "2002",
      pages          = "3363-3386",
      doi            = "10.1063/1.1471366",
      eprint         = "hep-ph/0110083",
      archivePrefix  = "arXiv",
      primaryClass   = "hep-ph",
      reportNumber   = "TTP-01-25, UPRF-2001-21",
      SLACcitation   = "
}

@article{Brown2004527,
      title = "Polylogarithmes multiples uniformes en une variable ",
      journal = "Comptes Rendus Mathematique ",
      volume = "338",
      number = "7",
      pages = "527 - 532",
      year = "2004",
      note = "",
      issn = "1631-073X",
      doi = "http://dx.doi.org/10.1016/j.crma.2004.02.001",
      author = "Francis C.S. Brown"
}

@article{Dixon:2014voa,
      author         = "Dixon, Lance J. and Drummond, James M. and Duhr, Claude
                        and Pennington, Jeffrey",
      title          = "{The four-loop remainder function and multi-Regge
                        behavior at NNLLA in planar N = 4 super-Yang-Mills
                        theory}",
      journal        = "JHEP",
      volume         = "06",
      year           = "2014",
      pages          = "116",
      doi            = "10.1007/JHEP06(2014)116",
      eprint         = "1402.3300",
      archivePrefix  = "arXiv",
      primaryClass   = "hep-th",
      reportNumber   = "SLAC-PUB-15902, CERN-PH-TH-2014-027, IPPP-14-09,
                        DCPT-14-18, LAPTH-010-14",
      SLACcitation   = "
}

@article{Bern:2005iz,
      author         = "Bern, Zvi and Dixon, Lance J. and Smirnov, Vladimir A.",
      title          = "{Iteration of planar amplitudes in maximally
                        supersymmetric Yang-Mills theory at three loops and
                        beyond}",
      journal        = "Phys. Rev.",
      volume         = "D72",
      year           = "2005",
      pages          = "085001",
      doi            = "10.1103/PhysRevD.72.085001",
      eprint         = "hep-th/0505205",
      archivePrefix  = "arXiv",
      primaryClass   = "hep-th",
      reportNumber   = "SLAC-PUB-11210, UCLA-05-TEP-14",
      SLACcitation   = "
}

@article{Basso:2014pla,
      author         = "Basso, Benjamin and Caron-Huot, Simon and Sever, Amit",
      title          = "{Adjoint BFKL at finite coupling: a short-cut from the
                        collinear limit}",
      journal        = "JHEP",
      volume         = "01",
      year           = "2015",
      pages          = "027",
      doi            = "10.1007/JHEP01(2015)027",
      eprint         = "1407.3766",
      archivePrefix  = "arXiv",
      primaryClass   = "hep-th",
      SLACcitation   = "
}

@article{Dixon:2012yy,
     author         = "Dixon, Lance J. and Duhr, Claude and Pennington, Jeffrey",
     title          = "{Single-valued harmonic polylogarithms and the
                       multi-Regge limit}",
     journal        = "JHEP",
     volume         = "1210",
     pages          = "074",
     doi            = "10.1007/JHEP10(2012)074",
     year           = "2012",
     eprint         = "1207.0186",
     archivePrefix  = "arXiv",
     primaryClass   = "hep-th",
     reportNumber   = "SLAC-PUB-15132",
     SLACcitation   = "
}

@article{Drummond:2015jea,
      author         = "Drummond, J. M. and Papathanasiou, G.",
      title          = "{Hexagon OPE Resummation and Multi-Regge Kinematics}",
      year           = "2015",
      eprint         = "1507.08982",
      archivePrefix  = "arXiv",
      primaryClass   = "hep-th",
      SLACcitation   = "
}

@article{Beisert:2006ez,
      author         = "Beisert, Niklas and Eden, Burkhard and Staudacher,
                        Matthias",
      title          = "{Transcendentality and Crossing}",
      journal        = "J. Stat. Mech.",
      volume         = "0701",
      year           = "2007",
      pages          = "P01021",
      doi            = "10.1088/1742-5468/2007/01/P01021",
      eprint         = "hep-th/0610251",
      archivePrefix  = "arXiv",
      primaryClass   = "hep-th",
      reportNumber   = "AEI-2006-079, ITP-UU-06-44, SPIN-06-34",
      SLACcitation   = "
}

@article{Pennington:2012zj,
      author         = "Pennington, Jeffrey",
      title          = "{The six-point remainder function to all loop orders in
                        the multi-Regge limit}",
      journal        = "JHEP",
      volume         = "01",
      year           = "2013",
      pages          = "059",
      doi            = "10.1007/JHEP01(2013)059",
      eprint         = "1209.5357",
      archivePrefix  = "arXiv",
      primaryClass   = "hep-th",
      reportNumber   = "SLAC-PUB-15251",
      SLACcitation   = "
}

@article{Apery,
     author         = "R.~Ap\'ery",
     title          = "{Irrationalit\'e de $\zm_2$ et $\z_3$}",
     journal        = "Ast\'erisque",
     volume         = "61",
     pages          = "11",
     year           = "1979",
}

@article{BallRivoal,
     author         = "K.~Ball and T.~Rivoal",
     title          = "{Irrationalit\'e d'une infinit\'e de valeurs de la fonction zeta aux entiers impairs.}",
     journal        = "Invent. Math.",
     volume         = "146",
     pages          = "193-207",
     year           = "2001",
}

@incollection {Zagier23,
             AUTHOR = {Zagier, Don},
              TITLE = {Values of zeta functions and their applications},
          BOOKTITLE = {First {E}uropean {C}ongress of {M}athematics, {V}ol.\ {II}
                       ({P}aris, 1992)},
             SERIES = {Progr. Math.},
             VOLUME = {120},
              PAGES = {497--512},
          PUBLISHER = {Birkh\"auser, Basel},
               YEAR = {1994},
            MRCLASS = {11M41 (11F67 11G40 19F27)},
           MRNUMBER = {1341859 (96k:11110)},
         MRREVIEWER = {Fernando Rodr{\'{\i}}guez Villegas},
}

@article{D'Hoker:2014gfa,
     author         = "D'Hoker, Eric and Green, Michael B. and Pioline, Boris
                       and Russo, Rodolfo",
     title          = "{Matching the $D^{6}R^{4}$ interaction at two-loops}",
     journal        = "JHEP",
     volume         = "1501",
     pages          = "031",
     doi            = "10.1007/JHEP01(2015)031",
     year           = "2015",
     eprint         = "1405.6226",
     archivePrefix  = "arXiv",
     primaryClass   = "hep-th",
     reportNumber   = "NSF-KITP-14045, DAMTP-2014-22, QMUL-PH-14-11,
                       CERN-PH-TH-2014-079",
     SLACcitation   = "
}

@article{Adams:2015gva,
     author         = "Adams, Luise and Bogner, Christian and Weinzierl, Stefan",
     title          = "{The two-loop sunrise integral around four space-time
                       dimensions and generalisations of the Clausen and Glaisher
                       functions towards the elliptic case}",
     year           = "2015",
     eprint         = "1504.03255",
     archivePrefix  = "arXiv",
     primaryClass   = "hep-ph",
     SLACcitation   = "
}

@proceedings{FECM,
     editor         = "A. Joseph et al.",
     title          = "Proc. of First European Congress of Mathematics (Paris 1992)",
     publisher      = {Birkh{\"{a}}user, Basel},
     year           = {1994},
}

@article{D'Hoker:2013eea,
     author         = "D'Hoker, Eric and Green, Michael B.",
     title          = "{Zhang-Kawazumi Invariants and Superstring Amplitudes}",
     year           = "2013",
     eprint         = "1308.4597",
     archivePrefix  = "arXiv",
     primaryClass   = "hep-th",
     reportNumber   = "DAMTP-2013-43",
     SLACcitation   = "
}

@article{LNT,
     author         = "Luque, Jean-Gabriel and Novelli, Jean-Christophe and Thibon, Jean-Yves",
     title          = "{Period polynomials and Ihara brackets}",
     year           = "2006",
     eprint         = "math/0606301",
     archivePrefix  = "arXiv",
     primaryClass   = "math.CO,math.NT",
}

@article{Broedel:2014vla,
      author         = "Broedel, Johannes and Mafra, Carlos R. and Matthes, Nils
                        and Schlotterer, Oliver",
      title          = "{Elliptic multiple zeta values and one-loop superstring
                        amplitudes}",
      journal        = "JHEP",
      volume         = "07",
      year           = "2015",
      pages          = "112",
      doi            = "10.1007/JHEP07(2015)112",
      eprint         = "1412.5535",
      archivePrefix  = "arXiv",
      primaryClass   = "hep-th",
      reportNumber   = "AEI-2014-066, DAMTP-2014-95",
      SLACcitation   = "
}

@unpublished{Pollack,
     author         = "Pollack, Aaron",
     title          = "{Relations between derivations arising from modular forms}",
     note           = "Undergraduate thesis, Duke University",
}

@incollection {KZB,
             AUTHOR = {Calaque, Damien and Enriquez, Benjamin and Etingof, Pavel},
              TITLE = {Universal {KZB} equations: the elliptic case},
          BOOKTITLE = {Algebra, arithmetic, and geometry: in honor of {Y}u. {I}.
                       {M}anin. {V}ol. {I}},
             SERIES = {Progr. Math.},
             VOLUME = {269},
              PAGES = {165--266},
          PUBLISHER = {Birkh\"auser Boston, Inc., Boston, MA},
               YEAR = {2009},
            MRCLASS = {32G34 (11F55 17B37 20C08 32C38)},
           MRNUMBER = {2641173 (2011k:32018)},
         MRREVIEWER = {Gwyn Bellamy},
                DOI = {10.1007/978-0-8176-4745-2_5},
                URL = {http://dx.doi.org/10.1007/978-0-8176-4745-2\_5},
}

@article{Hain,
     author         = "Hain, Richard",
     title          = "{Notes on the universal elliptic KZB equation}",
     eprint         = "1309.0580",
     archivePrefix  = "arXiv",
     primaryClass   = "math.AG",
}
 
@book{igusa,
     title          ={Theta Functions},
     author         ={Igusa, J.},
     year           ={1972},
     publisher      ={Springer}, 
}

@book{fay,
     title          ={Theta Functions on Riemann Surfaces},
     author         ={Fay, J.},
     year           ={1973},
     publisher      ={Springer}, 
}

@article{WIP,
     author         = "Broedel, Johannes and Mafra, Carlos and Matthes, Nils and Schlotterer, Oliver",
     title          = {},
     note           = {work in progress},
     year           = "2015",
}

@article{Levin,
     author         = "Levin, Andrey",
     title          = "{Elliptic polylogarithms: An analytic theory}",
     journal        = "Compositio Mathematica",
     volume         = "106",
     pages          = "267",
     year           = "1997",
}

@article{CaronHuot:2012ab,
     author         = "Caron-Huot, Simon and Larsen, Kasper J.",
     title          = "{Uniqueness of two-loop master contours}",
     journal        = "JHEP",
     volume         = "1210",
     pages          = "026",
     doi            = "10.1007/JHEP10(2012)026",
     year           = "2012",
     eprint         = "1205.0801",
     archivePrefix  = "arXiv",
     primaryClass   = "hep-ph",
     reportNumber   = "SACLAY-IPHT-T12-028, UUITP-13-12",
     SLACcitation   = "
}

@article{Bern:2008qj,
     author         = "Bern, Z. and Carrasco, J.J.M. and Johansson, Henrik",
     title          = "{New Relations for Gauge-Theory Amplitudes}",
     journal        = "Phys.Rev.",
     volume         = "D78",
     pages          = "085011",
     doi            = "10.1103/PhysRevD.78.085011",
     year           = "2008",
     eprint         = "0805.3993",
     archivePrefix  = "arXiv",
     primaryClass   = "hep-ph",
     reportNumber   = "UCLA-07-TEP-15",
     SLACcitation   = "
}

@article{Green:1999pv,
     author         = "Green, Michael B. and Vanhove, Pierre",
     title          = "{The Low-energy expansion of the one loop type II
                       superstring amplitude}",
     journal        = "Phys.Rev.",
     volume         = "D61",
     pages          = "104011",
     doi            = "10.1103/PhysRevD.61.104011",
     year           = "2000",
     eprint         = "hep-th/9910056",
     archivePrefix  = "arXiv",
     primaryClass   = "hep-th",
     reportNumber   = "DAMTP-1999-124, CERN-TH-99-200, SACLAY-SPH-T-99-071",
     SLACcitation   = "
}

@article{Broedel:2013tta,
     author         = "Broedel, Johannes and Schlotterer, Oliver and Stieberger,
                       Stephan",
     title          = "{Polylogarithms, Multiple Zeta Values and Superstring
                       Amplitudes}",
     journal        = "Fortsch.Phys.",
     volume         = "61",
     pages          = "812-870",
     doi            = "10.1002/prop.201300019",
     year           = "2013",
     eprint         = "1304.7267",
     archivePrefix  = "arXiv",
     primaryClass   = "hep-th",
     reportNumber   = "DAMTP-2013-22, AEI-2013-194, MPP-2013-119",
     SLACcitation   = "
}

@article{Mafra:2012kh,
     author         = "Mafra, Carlos R. and Schlotterer, Oliver",
     title          = "{The Structure of n-Point One-Loop Open Superstring
                       Amplitudes}",
     journal        = "JHEP",
     volume         = "1408",
     pages          = "099",
     doi            = "10.1007/JHEP08(2014)099",
     year           = "2014",
     eprint         = "1203.6215",
     archivePrefix  = "arXiv",
     primaryClass   = "hep-th",
     reportNumber   = "AEI-2012-032, DAMTP-2012-26",
     SLACcitation   = "
} 

@article{Schlotterer:2012ny,
     author         = "Schlotterer, Oliver and Stieberger, Stephan",
     title          = "{Motivic Multiple Zeta Values and Superstring
                       Amplitudes}",
     journal        = "J.Phys.",
     volume         = "A46",
     pages          = "475401",
     doi            = "10.1088/1751-8113/46/47/475401",
     year           = "2013",
     eprint         = "1205.1516",
     archivePrefix  = "arXiv",
     primaryClass   = "hep-th",
     reportNumber   = "AEI-2012-039, MPP-2012-859",
     SLACcitation   = "
}

@article{D'Hoker:2015foa,
     author         = "D'Hoker, Eric and Green, Michael B. and Vanhove, Pierre",
     title          = "{On the modular structure of the genus-one Type II
                       superstring low energy expansion}",
     year           = "2015",
     eprint         = "1502.06698",
     archivePrefix  = "arXiv",
     primaryClass   = "hep-th",
     reportNumber   = "DAMTP-08-02-2015, IPHT-T15-012, IHES-P-15-04",
     SLACcitation   = "
}

@article{BrownLev,
     author         = "Brown, Francis. and Levin, Andrey.",
     title          = "{Multiple elliptic polylogarithms}",
     year           = "2011",
     eprint         = "arXiv:1110.6917v2",
     archivePrefix  = "arXiv",
     primaryClass   = "math",
}

@article{Tsuchiya:1988va,
     author         = "Tsuchiya, Akihiko",
     title          = "{More on One Loop Massless Amplitudes of Superstring
                       Theories}",
     journal        = "Phys.Rev.",
     volume         = "D39",
     pages          = "1626",
     doi            = "10.1103/PhysRevD.39.1626",
     year           = "1989",
     reportNumber   = "TIT/HEP-135",
     SLACcitation   = "
}

@article{Barreiro:2012aw,
     author         = "Barreiro, Luiz Antonio and Medina, Ricardo",
     title          = "{Revisiting the S-matrix approach to the open superstring
                       low energy effective lagrangian}",
     journal        = "JHEP",
     volume         = "1210",
     pages          = "108",
     doi            = "10.1007/JHEP10(2012)108",
     year           = "2012",
     eprint         = "1208.6066",
     archivePrefix  = "arXiv",
     primaryClass   = "hep-th",
     SLACcitation   = "
}

@book{Green:1987mn,
     author         = "Green, Michael B. and Schwarz, J.H. and Witten, Edward",
     title          = "{Superstring Theory. Vol. 2: Loop amplitudes, anomalies and phenomenology}",
     year           = "1987",
     publisher      = "Cambridge, UK: Univ. Pr. (1987) (Cambridge Monographs on Mathematical Physics)", 
     SLACcitation   = "
}

@book{Green:1987sp,
     author         = "Green, Michael B. and Schwarz, J.H. and Witten, Edward",
     title          = "{Superstring Theory. Vol. 1: Introduction}",
     year           = "1987",
     publisher      = "Cambridge, UK: Univ. Pr. (1987) (Cambridge Monographs on Mathematical Physics)", 
     SLACcitation   = "
}

@book{Weil:1976,
     author         = "Weil, Andr\'e",
     title          = "{Elliptic functions according to Eisenstein and Kronecker}",
     year           = "1976",
     publisher      = "Springer, Heidelberg, Published in ``Ergebnisse der Mathematik und ihrer Grenzgebiete''", 
}

@book{Cameron,
     author         = "Peter J. Cameron",
     title          = "{Combinatorics. Topics, techniques, algorithms}",
     year           = "1994",
     publisher      = "Cambridge, Uk: Univ. Pr.",
}

@book{Riordan,
     title          = {Introduction to Combinatorial Analysis},
     author         = {John Riordan},
     publisher      = {Dover Publications},
     isbn           = {9780486425368,0486425363},
     year           = {2002},
     series         = {},
     edition        = {},
     volume         = {},
}

@book{StanleyBook,
     author         = "Stanley, Richard P.",
     title          = "{Enumerative Combinatorics}",
     year           = "2012",
     publisher      = "Cambridge, UK: Univ. Pr.",
     volume         = {1},
     edition        = {second edition},
}

@url{oeis,
     url            = "http://oeis.org",
}

@url{MZVWebsite,
     url            = "http://mzv.mpp.mpg.de",
}

@article{Stieberger:2009hq,
     author         = "Stieberger, S.",
     title          = "{Open &amp; Closed vs. Pure Open String Disk Amplitudes}",
     year           = "2009",
     eprint         = "0907.2211",
     archivePrefix  = "arXiv",
     primaryClass   = "hep-th",
     reportNumber   = "MPP-2008-01",
     SLACcitation   = "
}

@article{BrownTate,
     author         = "Brown, F.",
     title          = "{Mixed Tate motives over $\ZZ$}",
     journal        = "Ann. Math.",
     volume         = "175",
     pages          = "949",
     year           = "2012",
}

@article{Goncharov:2001iea,
     author         = "Goncharov, A.B.",
     title          = "{Multiple polylogarithms and mixed Tate motives}",
     year           = "2001",
     eprint         = "math/0103059",
     archivePrefix  = "arXiv",
     primaryClass   = "math.AG",
     SLACcitation   = "
}

@article {Goncharov:2005sla,
             AUTHOR = {Goncharov, A. B.},
              TITLE = {Galois symmetries of fundamental groupoids and noncommutative
                       geometry},
            JOURNAL = {Duke Math. J.},
           FJOURNAL = {Duke Mathematical Journal},
             VOLUME = {128},
               YEAR = {2005},
             NUMBER = {2},
              PAGES = {209--284},
               ISSN = {0012-7094},
              CODEN = {DUMJAO},
            MRCLASS = {11G55 (11G09 14C30 16W30 19E15 20F34)},
           MRNUMBER = {2140264 (2007b:11094)},
         MRREVIEWER = {Matilde Marcolli},
                DOI = {10.1215/S0012-7094-04-12822-2},
                URL = {http://dx.doi.org/10.1215/S0012-7094-04-12822-2},
}

@article{Remiddi:1999ew,
     author         = "Remiddi, E. and Vermaseren, J.A.M.",
     title          = "{Harmonic polylogarithms}",
     journal        = "Int.J.Mod.Phys.",
     volume         = "A15",
     pages          = "725-754",
     doi            = "10.1142/S0217751X00000367",
     year           = "2000",
     eprint         = "hep-ph/9905237",
     archivePrefix  = "arXiv",
     primaryClass   = "hep-ph",
     reportNumber   = "NIKHEF-99-005, TTP-99-08",
     SLACcitation   = "
}

@article{Duhr:2011zq,
     author         = "Duhr, Claude and Gangl, Herbert and Rhodes, John R.",
     title          = "{From polygons and symbols to polylogarithmic functions}",
     journal        = "JHEP",
     volume         = "1210",
     pages          = "075",
     doi            = "10.1007/JHEP10(2012)075",
     year           = "2012",
     eprint         = "1110.0458",
     archivePrefix  = "arXiv",
     primaryClass   = "math-ph",
     reportNumber   = "IPPP-11-56, DCPT-11-112",
     SLACcitation   = "
}

@article{Goncharov.A.B.:2009tja,
     author         = "Goncharov, A.B.",
     title          = "{A simple construction of Grassmannian polylogarithms}",
     year           = "2009",
     eprint         = "0908.2238",
     archivePrefix  = "arXiv",
     primaryClass   = "math.AG",
     SLACcitation   = "
}

@article{Maitre:2005uu,
     author         = "Maitre, D",
     title          = "{HPL, a mathematica implementation of the harmonic
                       polylogarithms}",
     journal        = "Comput.Phys.Commun.",
     volume         = "174",
     pages          = "222-240",
     doi            = "10.1016/j.cpc.2005.10.008",
     year           = "2006",
     eprint         = "hep-ph/0507152",
     archivePrefix  = "arXiv",
     primaryClass   = "hep-ph",
     reportNumber   = "ZU-TH-14-05",
     SLACcitation   = "
}

@article{Goncharov:2010jf,
     author         = "Goncharov, Alexander B. and Spradlin, Marcus and Vergu,
                       C. and Volovich, Anastasia",
     title          = "{Classical Polylogarithms for Amplitudes and Wilson
                       Loops}",
     journal        = "Phys.Rev.Lett.",
     volume         = "105",
     pages          = "151605",
     doi            = "10.1103/PhysRevLett.105.151605",
     year           = "2010",
     eprint         = "1006.5703",
     archivePrefix  = "arXiv",
     primaryClass   = "hep-th",
     reportNumber   = "BROWN-HET-1602",
     SLACcitation   = "
}

@article{Broedel:2013aza,
     author         = "Broedel, Johannes and Schlotterer, Oliver and Stieberger,
                       Stephan and Terasoma, Tomohide",
     title          = "{All order $\alpha^{\prime}$-expansion of superstring
                       trees from the Drinfeld associator}",
     journal        = "Phys.Rev.",
     number         = "6",
     volume         = "D89",
     pages          = "066014",
     doi            = "10.1103/PhysRevD.89.066014",
     year           = "2014",
     eprint         = "1304.7304",
     archivePrefix  = "arXiv",
     primaryClass   = "hep-th",
     reportNumber   = "DAMTP-2013-23, AEI-2013-195, MPP-2013-120",
     SLACcitation   = "
}

@article{Tsuchiya:2012nf,
     author         = "Tsuchiya, A.G.",
     title          = "{On the pole structures of the disconnected part of hyperelliptic g-loop M-point super string amplitudes}",
     year           = "2012",
     eprint         = "1209.6117",
     archivePrefix  = "arXiv",
     primaryClass   = "hep-th",
     SLACcitation   = "
}

@article{Mafra:2011nv,
     author         = "Mafra, Carlos R. and Schlotterer, Oliver and Stieberger,
                       Stephan",
     title          = "{Complete N-Point Superstring Disk Amplitude I. Pure
                       Spinor Computation}",
     journal        = "Nucl.Phys.",
     volume         = "B873",
     pages          = "419-460",
     doi            = "10.1016/j.nuclphysb.2013.04.023",
     year           = "2013",
     eprint         = "1106.2645",
     archivePrefix  = "arXiv",
     primaryClass   = "hep-th",
     reportNumber   = "AEI-2011-34, MPP-2011-47",
     SLACcitation   = "
}

@article{Mafra:2011nw,
     author         = "Mafra, Carlos R. and Schlotterer, Oliver and Stieberger,
                       Stephan",
     title          = "{Complete N-Point Superstring Disk Amplitude II.
                       Amplitude and Hypergeometric Function Structure}",
     journal        = "Nucl.Phys.",
     volume         = "B873",
     pages          = "461-513",
     doi            = "10.1016/j.nuclphysb.2013.04.022",
     year           = "2013",
     eprint         = "1106.2646",
     archivePrefix  = "arXiv",
     primaryClass   = "hep-th",
     reportNumber   = "AEI-2011-35, MPP-2011-65",
     SLACcitation   = "
}

@article{Oprisa:2005wu,
     author         = "Oprisa, D. and Stieberger, S.",
     title          = "{Six gluon open superstring disk amplitude, multiple
                       hypergeometric series and Euler-Zagier sums}",
     year           = "2005",
     eprint         = "hep-th/0509042",
     archivePrefix  = "arXiv",
     primaryClass   = "hep-th",
     reportNumber   = "LMU-ASC-07-05, MPP-2005-12",
     SLACcitation   = "
}

@article{Green:1981ya,
     author         = "Green, Michael B. and Schwarz, John H.",
     title          = "{Supersymmetrical Dual String Theory. 3. Loops and
                       Renormalization}",
     journal        = "Nucl.Phys.",
     volume         = "B198",
     pages          = "441-460",
     doi            = "10.1016/0550-3213(82)90334-0",
     year           = "1982",
     reportNumber   = "CALT-68-873",
     SLACcitation   = "
}

@article{Schwarz:1982jn,
     author         = "Schwarz, John H.",
     title          = "{Superstring Theory}",
     journal        = "Phys.Rept.",
     volume         = "89",
     pages          = "223-322",
     doi            = "10.1016/0370-1573(82)90087-4",
     year           = "1982",
     reportNumber   = "CALT-68-911",
     SLACcitation   = "
}

@article{Green:1984ed,
     author         = "Green, Michael B. and Schwarz, John H.",
     title          = "{Infinity Cancellations in SO(32) Superstring Theory}",
     journal        = "Phys.Lett.",
     volume         = "B151",
     pages          = "21-25",
     doi            = "10.1016/0370-2693(85)90816-0",
     year           = "1985",
     reportNumber   = "CALT-68-1194",
     SLACcitation   = "
}

@article{Green:1984sg,
     author         = "Green, Michael B. and Schwarz, John H.",
     title          = "{Anomaly Cancellation in Supersymmetric D=10 Gauge Theory
                       and Superstring Theory}",
     journal        = "Phys.Lett.",
     volume         = "B149",
     pages          = "117-122",
     doi            = "10.1016/0370-2693(84)91565-X",
     year           = "1984",
     reportNumber   = "CALT-68-1182",
     SLACcitation   = "
}

@article{Green:1984qs,
     author         = "Green, Michael B. and Schwarz, John H.",
     title          = "{The Hexagon Gauge Anomaly in Type I Superstring Theory}",
     journal        = "Nucl.Phys.",
     volume         = "B255",
     pages          = "93-114",
     doi            = "10.1016/0550-3213(85)90130-0",
     year           = "1985",
     reportNumber   = "CALT-68-1224",
     SLACcitation   = "
}

@article{Green:1982sw,
     author         = "Green, Michael B. and Schwarz, John H. and Brink, Lars",
     title          = "{N=4 Yang-Mills and N=8 Supergravity as Limits of String
                       Theories}",
     journal        = "Nucl.Phys.",
     volume         = "B198",
     pages          = "474-492",
     doi            = "10.1016/0550-3213(82)90336-4",
     year           = "1982",
     reportNumber   = "CALT-68-880",
     SLACcitation   = "
}

@article{Green:2013bza,
     author         = "Green, Michael B. and Mafra, Carlos R. and Schlotterer,
                       Oliver",
     title          = "{Multiparticle one-loop amplitudes and S-duality in
                       closed superstring theory}",
     journal        = "JHEP",
     volume         = "1310",
     pages          = "188",
     doi            = "10.1007/JHEP10(2013)188",
     year           = "2013",
     eprint         = "1307.3534",
     archivePrefix  = "arXiv",
     reportNumber   = "AEI-2013-219, DAMTP-2013-33",
     SLACcitation   = "
}

@article{Richards:2008jg,
     author         = "Richards, David M.",
     title          = "{The One-Loop Five-Graviton Amplitude and the Effective
                       Action}",
     journal        = "JHEP",
     volume         = "0810",
     pages          = "042",
     doi            = "10.1088/1126-6708/2008/10/042",
     year           = "2008",
     eprint         = "0807.2421",
     archivePrefix  = "arXiv",
     primaryClass   = "hep-th",
     reportNumber   = "DAMTP-2008-60",
     SLACcitation   = "
}

@article{Green:2008uj,
     author         = "Green, Michael B. and Russo, Jorge G. and Vanhove,
                       Pierre",
     title          = "{Low energy expansion of the four-particle genus-one
                       amplitude in type II superstring theory}",
     journal        = "JHEP",
     volume         = "0802",
     pages          = "020",
     doi            = "10.1088/1126-6708/2008/02/020",
     year           = "2008",
     eprint         = "0801.0322",
     archivePrefix  = "arXiv",
     primaryClass   = "hep-th",
     reportNumber   = "DAMTP-2007-96, SPHT-T-07-126, UB-ECM-PF-07-29",
     SLACcitation   = "
}

@article{Ramond:1971gb,
     author         = "Ramond, Pierre",
     title          = "{Dual Theory for Free Fermions}",
     journal        = "Phys.Rev.",
     volume         = "D3",
     pages          = "2415-2418",
     doi            = "10.1103/PhysRevD.3.2415",
     year           = "1971",
     reportNumber   = "FERMILAB-PUB-70-008-T, FERMILAB-PUB-70-008-THY, THY-8",
     SLACcitation   = "
}

@article{Neveu:1971rx,
     author         = "Neveu, A. and Schwarz, J.H.",
     title          = "{Factorizable dual model of pions}",
     journal        = "Nucl.Phys.",
     volume         = "B31",
     pages          = "86-112",
     doi            = "10.1016/0550-3213(71)90448-2",
     year           = "1971",
     SLACcitation   = "
}

@article{Neveu:1971iv,
     author         = "Neveu, A. and Schwarz, J.H.",
     title          = "{Quark model of dual pions}",
     journal        = "Phys.Rev.",
     volume         = "D4",
     pages          = "1109-1111",
     doi            = "10.1103/PhysRevD.4.1109",
     year           = "1971",
     SLACcitation   = "
}

@article{Brown:2013gia,
     author         = "Brown, Francis",
     title          = "{Single-valued periods and multiple zeta values}",
     year           = "2013",
     eprint         = "1309.5309",
     archivePrefix  = "arXiv",
     primaryClass   = "math.NT",
     SLACcitation   = "
}

@article{Drummond:2013vz,
     author         = "Drummond, J.M. and Ragoucy, E.",
     title          = "{Superstring amplitudes and the associator}",
     journal        = "JHEP",
     volume         = "1308",
     pages          = "135",
     doi            = "10.1007/JHEP08(2013)135",
     year           = "2013",
     eprint         = "1301.0794",
     archivePrefix  = "arXiv",
     primaryClass   = "hep-th",
     reportNumber   = "CERN-PH-TH-2012-363, LAPTH-062-12",
     SLACcitation   = "
}

@article{Stieberger:2009rr,
     author         = "Stieberger, Stephan",
     title          = "{Constraints on Tree-Level Higher Order Gravitational
                       Couplings in Superstring Theory}",
     journal        = "Phys.Rev.Lett.",
     volume         = "106",
     pages          = "111601",
     doi            = "10.1103/PhysRevLett.106.111601",
     year           = "2011",
     eprint         = "0910.0180",
     archivePrefix  = "arXiv",
     primaryClass   = "hep-th",
     reportNumber   = "MPP-2009-158",
     SLACcitation   = "
}

@article{Stieberger:2013wea,
     author         = "Stieberger, S.",
     title          = "{Closed superstring amplitudes, single-valued multiple
                       zeta values and the Deligne associator}",
     journal        = "J.Phys.",
     volume         = "A47",
     pages          = "155401",
     doi            = "10.1088/1751-8113/47/15/155401",
     year           = "2014",
     eprint         = "1310.3259",
     archivePrefix  = "arXiv",
     primaryClass   = "hep-th",
     reportNumber   = "MPP-2013-278",
     SLACcitation   = "
}

@article{Stieberger:2014hba,
     author         = "Stieberger, Stephan and Taylor, Tomasz R.",
     title          = "{Closed String Amplitudes as Single-Valued Open String
                       Amplitudes}",
     journal        = "Nucl.Phys.",
     volume         = "B881",
     pages          = "269-287",
     doi            = "10.1016/j.nuclphysb.2014.02.005",
     year           = "2014",
     eprint         = "1401.1218",
     archivePrefix  = "arXiv",
     primaryClass   = "hep-th",
     reportNumber   = "MPP-2014-001",
     SLACcitation   = "
}

@article{Adams:2014vja,
     author         = "Adams, Luise and Bogner, Christian and Weinzierl, Stefan",
     title          = "{The two-loop sunrise graph in two space-time dimensions
                       with arbitrary masses in terms of elliptic dilogarithms}",
     journal        = "J.Math.Phys.",
     number         = "10",
     volume         = "55",
     pages          = "102301",
     doi            = "10.1063/1.4896563",
     year           = "2014",
     eprint         = "1405.5640",
     archivePrefix  = "arXiv",
     primaryClass   = "hep-ph",
     SLACcitation   = "
}

@article{Bloch:2013tra,
    author         = "Bloch, Spencer and Vanhove, Pierre",
    title          = "{The elliptic dilogarithm for the sunset graph}",
    journal        = "J. Number Theory",
    volume         = "148",
    pages          = "328--364",
    year           = "2015",
    eprint         = "1309.5865",
    archivePrefix  = "arXiv",
    primaryClass   = "hep-th",
    SLACcitation   = "
}

@article{Bloch:2014qca,
     author         = "Bloch, Spencer and Kerr, Matt and Vanhove, Pierre",
     title          = "{A Feynman integral via higher normal functions}",
     year           = "2014",
     eprint         = "1406.2664",
     archivePrefix  = "arXiv",
     primaryClass   = "hep-th",
     reportNumber   = "IPHT-T-14-015, IHES-P-14-06",
     SLACcitation   = "
}

@incollection {Brown:2011ik,
             AUTHOR = {Brown, Francis C. S.},
              TITLE = {On the decomposition of motivic multiple zeta values},
          BOOKTITLE = {Galois-{T}eichm\"uller theory and arithmetic geometry},
             SERIES = {Adv. Stud. Pure Math.},
             VOLUME = {63},
              PAGES = {31--58},
          PUBLISHER = {Math. Soc. Japan, Tokyo},
               YEAR = {2012},
     eprint         = "1102.1310",
     archivePrefix  = "arXiv",
     primaryClass   = "math.NT",
     SLACcitation   = "
}

@article{Duhr:2012fh,
     author         = "Duhr, Claude",
     title          = "{Hopf algebras, coproducts and symbols: an application to
                       Higgs boson amplitudes}",
     journal        = "JHEP",
     volume         = "1208",
     pages          = "043",
     doi            = "10.1007/JHEP08(2012)043",
     year           = "2012",
     eprint         = "1203.0454",
     archivePrefix  = "arXiv",
     primaryClass   = "hep-ph",
     SLACcitation   = "
}

@article{Schnetz:2013hqa,
     author         = "Schnetz, Oliver",
     title          = "{Graphical functions and single-valued multiple
                       polylogarithms}",
     year           = "2013",
     eprint         = "1302.6445",
     archivePrefix  = "arXiv",
     primaryClass   = "math.NT",
     SLACcitation   = "
}

@article{Stieberger:2006te,
     author         = "Stieberger, Stephan and Taylor, Tomasz R.",
     title          = "{Multi-Gluon Scattering in Open Superstring Theory}",
     journal        = "Phys.Rev.",
     volume         = "D74",
     pages          = "126007",
     doi            = "10.1103/PhysRevD.74.126007",
     year           = "2006",
     eprint         = "hep-th/0609175",
     archivePrefix  = "arXiv",
     primaryClass   = "hep-th",
     reportNumber   = "LMU-ASC-50-06",
     SLACcitation   = "
}

@article{DelDuca:2013lma,
     author         = "Del Duca, Vittorio and Dixon, Lance J. and Duhr, Claude
                       and Pennington, Jeffrey",
     title          = "{The BFKL equation, Mueller-Navelet jets and
                       single-valued harmonic polylogarithms}",
     journal        = "JHEP",
     volume         = "1402",
     pages          = "086",
     doi            = "10.1007/JHEP02(2014)086",
     year           = "2014",
     eprint         = "1309.6647",
     archivePrefix  = "arXiv",
     primaryClass   = "hep-ph",
     reportNumber   = "SLAC-PUB-15706, IPPP-13-76, DCPT-13-152",
     SLACcitation   = "
}

@article{LevinRacinet,
     author         = "A. Levin and G. Racinet",
     title          = "{Towards multiple elliptic polylogarithms}",
     year           = "2007",
     eprint         = "arxiv:math/0703237",
     archivePrefix  = "arXiv",
     primaryClass   = "math",
}

@inproceedings{BeilinsonLevin,
     author         = "Beilinson, A.A. and Levin, A.",
     title          = "{The Elliptic Polylogarithm}",
     crossref       ={motives},
     pages          ={123-190},
}

@proceedings{motives,
     editor         ={U. Jannsen, S.L. Kleiman, J.-P. Serre},
     title          ="Motives",
     booktitle      ="Proc. of Symp. in Pure Math. 55, Part II",
     publisher      ={AMS},
     year           ={1994},
}

@inproceedings{Brown:ICM14,
     author         ={F. Brown},
     eprint         = "1407.5165",
     crossref       ={ICM14},
     title          ="Motivic Periods and the Projective Line minus Three Points",
}

@proceedings{ICM14,
     booktitle      ="Proceedings of the ICM 2014",
}

@book{Wildeshaus,
     title          ={Realizations of Polylogarithms},
     author         ={J\"org Wildeshaus},
     publisher      = {Springer},
     isbn           ={3540624600,9783540624608},
     year           ={1997},
     series         ={Lecture Notes in Mathematics},
     volume         ={1650},
}

@article{EnriquezEMZV,
      hyphenation   = {american},
      author        = {Benjamin Enriquez},
      title         = {Analogues elliptiques des nombres multiz{\'e}tas},
      date          = {2013-01-14},
      year          = {2013},
      eprinttype    = {arxiv},
      archivePrefix = {arXiv},
      eprint        = {1301.3042}
} 

@article{Adams:2013nia,
     author         = "Adams, Luise and Bogner, Christian and Weinzierl, Stefan",
     title          = "{The two-loop sunrise graph with arbitrary masses}",
     journal        = "J.Math.Phys.",
     volume         = "54",
     pages          = "052303",
     doi            = "10.1063/1.4804996",
     year           = "2013",
     eprint         = "1302.7004",
     archivePrefix  = "arXiv",
     primaryClass   = "hep-ph",
     SLACcitation   = "
}

@article{Clavelli:1986fj,
     author         = "Clavelli, L. and Cox, Paul H. and Harms, B.",
     title          = "{Parity Violating One Loop Six Point Function in Type I
                       Superstring Theory}",
     journal        = "Phys.Rev.",
     volume         = "D35",
     pages          = "1908",
     doi            = "10.1103/PhysRevD.35.1908",
     year           = "1987",
     reportNumber   = "UA-HEP-861, C86-07-16",
     SLACcitation   = "
}

@article{Mafra:2014gsa,
     author         = "Mafra, Carlos R. and Schlotterer, Oliver",
     title          = "{Cohomology foundations of one-loop amplitudes in pure
                       spinor superspace}",
     year           = "2014",
     eprint         = "1408.3605",
     archivePrefix  = "arXiv",
     primaryClass   = "hep-th",
     SLACcitation   = "
}

@article{Berkovits:2000fe,
     author         = "Berkovits, Nathan",
     title          = "{Super Poincare covariant quantization of the
                       superstring}",
     journal        = "JHEP",
     volume         = "0004",
     pages          = "018",
     doi            = "10.1088/1126-6708/2000/04/018",
     year           = "2000",
     eprint         = "hep-th/0001035",
     archivePrefix  = "arXiv",
     primaryClass   = "hep-th",
     reportNumber   = "IFT-P-005-2000",
     SLACcitation   = "
}

@book{Bloch:2000,
     author         = {Bloch, Spencer J.},
     isbn           = {0-8218-2114-8},
     mrclass        = {11G55 (11G40 11R70 14G10 19F27)},
     mrnumber       = {1760901 (2001i:11082)},
     mrreviewer     = {Jan Nekov{\'a}{\v{r}}},
     pages          = {1-97},
     publisher      = {American Mathematical Society, Providence, RI},
     series         = {CRM Monograph Series},
     title          = {Higher regulators, algebraic {$K$}-theory, and zeta functions of elliptic curves},
     volume         = {11},
     year           = {2000},
}

@article{Bogner:2014mha,
     author         = "Bogner, Christian and Brown, Francis",
     title          = "{Feynman integrals and iterated integrals on moduli
                       spaces of curves of genus zero}",
     year           = "2014",
     eprint         = "1408.1862",
     archivePrefix  = "arXiv",
     primaryClass   = "hep-th",
     SLACcitation   = "
}

@article{Brown:mmv,
     author         = {Brown, Francis},
     title          = {Multiple modular values for $\SL_2(\ZZ)$},
     year           = {2014},
     eprint         = "1407.5167v1",
     archivePrefix  = "arXiv",
     primaryClass   = "math.NT",
}

@article {Brown:0606,
             AUTHOR = {Brown, Francis C. S.},
              TITLE = {Multiple zeta values and periods of moduli spaces
                       $\overline{\germ M}_{0,n}$},
            JOURNAL = {Ann. Sci. \'Ec. Norm. Sup\'er. (4)},
           FJOURNAL = {Annales Scientifiques de l'\'Ecole Normale Sup\'erieure.
                       Quatri\`eme S\'erie},
             VOLUME = {42},
               YEAR = {2009},
             NUMBER = {3},
              PAGES = {371--489},
               ISSN = {0012-9593},
            MRCLASS = {32G20 (11G55 11M32 14F42 14H10 19F27 33B30)},
           MRNUMBER = {2543329 (2010f:32013)},
         MRREVIEWER = {Wadim Zudilin},
}

@article{Brown:kthree,
     author         = {Brown, Francis and Schnetz, Oliver},
     journal        = "Duke Math.J.",
     number         = {10},
     pages          = {1817--1862},
     title          = {A K3 in $\phi^4$},
     volume         = {161},
     year           = {2012},
}

@article{Brown:2013wn,
     author         = "Brown, Francis and Doryn, Dzmitry",
     title          = "{Framings for graph hypersurfaces}",
     year           = "2013",
     eprint         = "1301.3056",
     archivePrefix  = "arXiv",
     primaryClass   = "math.AG",
     SLACcitation   = "
}

@article{EnriquezEllAss,
     author         = {Enriquez, Benjamin},
     doi            = {10.1007/s00029-013-0137-3},
     fjournal       = {Selecta Mathematica. New Series},
     issn           = {1022-1824},
     journal        = {Selecta Math. (N.S.)},
     mrclass        = {17B35 (11M32 14H10 16S30 20F36)},
     mrnumber       = {3177926},
     number         = {2},
     pages          = {491--584},
     title          = {Elliptic associators},
     url            = {http://dx.doi.org/10.1007/s00029-013-0137-3},
     volume         = {20},
     year           = {2014},
}

@unpublished{Matthes:edzv,
     author         = {Matthes, Nils},
     note           = {in preparation},
     title          = {Elliptic double zeta values},
}

@unpublished{Matthes:thesis,
     author         = {Matthes, Nils},
     note           = {work in progress},
     title          = {},
}

@book{mumford1984tata,
     author         = {Mumford, D. and Nori, M. and Norman, P.},
     publisher      = {Birkh{\"a}user},
     title          = {Tata Lectures on Theta I, II},
     year           = {1983, 1984},
     isbn           = {9780817631109},
     number         = {Bd. 2},
     url            = {http://books.google.com/books?id=jPFaLY31gwYC},
}

@article{Zagier,
     author         = {Zagier, Don},
     coden          = {MAANA},
     doi            = {10.1007/BF01453591},
     fjournal       = {Mathematische Annalen},
     issn           = {0025-5831},
     journal        = {Math. Ann.},
     mrclass        = {11R42 (11R70 19F27)},
     mrnumber       = {1032949 (90k:11153)},
     mrreviewer     = {V. Kumar Murty},
     number         = {1-3},
     pages          = {613--624},
     title          = {The {B}loch-{W}igner-{R}amakrishnan polylogarithm function},
     url            = {http://dx.doi.org/10.1007/BF01453591},
     volume         = {286},
     year           = {1990},
}

@article{ZagierF,
     author         = "Zagier, D",
     title          = "{Periods of modular forms and Jacobi theta functions}",
     journal        = "Invent. Math.",
     volume         = "104",
     pages          = "449-465",
     year           = "1991",
}

@article{Kronecker,
     author         = "Kronecker, L.",
     title          = "{Zur Theorie der elliptischen Funktionen}",
     journal        = "Mathematische Werke",
     volume         = "IV",
     pages          = "313-318",
     year           = "1881",
}

@article{Namazie:1986zu,
     author         = "Namazie, M.A. and Narain, K.S. and Sarmadi, M.H.",
     title          = "{On Loop Amplitudes in the Fermionic String}",
     year           = "1986",
     reportNumber   = "RAL-86-051",
     SLACcitation   = "
}

@article{Le,
     author         = "Le, T.Q.T. and Murakami, J.",
     title          = "{Kontsevich's integral for the Kauffman polynomial}",
     journal        = "Nagoya Math J.",
     volume         = "142",
     pages          = "93-65",
     year           = "1996",
}

@preamble{
  "\def\cprime{$'$} "
}
@article {Drinfeld:1989st,
             AUTHOR = {Drinfeld, V. G.},
              TITLE = {Quasi-{H}opf algebras},
            JOURNAL = {Algebra i Analiz},
           FJOURNAL = {Algebra i Analiz},
             VOLUME = {1},
               YEAR = {1989},
             NUMBER = {6},
              PAGES = {114--148},
               ISSN = {0234-0852},
            MRCLASS = {17B37 (16W30 57M25 81T40)},
           MRNUMBER = {1047964 (91b:17016)},
         MRREVIEWER = {Ya. S. So{\u\i}bel{\cprime}man},
}

@article{Drinfeld2,
     author         = "Drinfeld, V.G.",
     title          = "{On quasitriangular quasi-Hopf algebras and on a group that is closely connected with $\text{Gal}(\bar{\mathbb Q}/ \mathbb Q)$}",
     journal        = "Leningrad Math.",
     volume         = "J. 2 (4)",
     pages          = "829-860",
     year           = "1991",
     SLACcitation   = "
}

@article{Terasoma,
    author          = "Terasoma, Tomohide",
    title           = "{Selberg Integrals and Multiple Zeta Values}",
    journal         = "Compositio Mathematica",
    volume          = "133",
    year            = "2002",
    pages           = "1--24",
}

@article{Stieberger:2002fh,
     author         = "Stieberger, S. and Taylor, T.R.",
     title          = "{NonAbelian Born-Infeld action and type I. Heterotic
                       duality (1): Heterotic F**6 terms at two loops}",
     journal        = "Nucl.Phys.",
     volume         = "B647",
     pages          = "49-68",
     doi            = "10.1016/S0550-3213(02)00895-7",
     year           = "2002",
     eprint         = "hep-th/0207026",
     archivePrefix  = "arXiv",
     primaryClass   = "hep-th",
     reportNumber   = "HUB-EP-02-26, NUB-3230",
     SLACcitation   = "
}

@article{Ablinger:2013cf,
     author         = "Ablinger, Jakob and Bluemlein, Johannes and Schneider,
                       Carsten",
     title          = "{Analytic and Algorithmic Aspects of Generalized Harmonic
                       Sums and Polylogarithms}",
     journal        = "J.Math.Phys.",
     volume         = "54",
     pages          = "082301",
     doi            = "10.1063/1.4811117",
     year           = "2013",
     eprint         = "1302.0378",
     archivePrefix  = "arXiv",
     primaryClass   = "math-ph",
     reportNumber   = "DESY-12-210, DO-TH-13-01, SFB-CPP-12-91, LPN-12-125",
     SLACcitation   = "
}

@article{Blumlein:2009cf,
     author         = "Blumlein, J. and Broadhurst, D.J. and Vermaseren, J.A.M.",
     title          = "{The Multiple Zeta Value Data Mine}",
     journal        = "Comput.Phys.Commun.",
     volume         = "181",
     pages          = "582-625",
     doi            = "10.1016/j.cpc.2009.11.007",
     year           = "2010",
     eprint         = "0907.2557",
     archivePrefix  = "arXiv",
     primaryClass   = "math-ph",
     reportNumber   = "DESY-09-003, SFB-CPP-09-65",
     SLACcitation   = "
}

@article{Ablinger:2013jta,
     author         = "Ablinger, Jakob and Bluemlein, Johannes",
     title          = "{Harmonic Sums, Polylogarithms, Special Numbers, and
                       their Generalizations}",
     year           = "2013",
     eprint         = "1304.7071",
     archivePrefix  = "arXiv",
     primaryClass   = "math-ph",
     reportNumber   = "DESY-13-073, DO-TH-13-10, SFB-CPP-13-26, LPN13-025",
     SLACcitation   = "
}

@article{Stieberger:2002wk,
     author         = "Stieberger, S. and Taylor, T.R.",
     title          = "{NonAbelian Born-Infeld action and type 1. - heterotic
                       duality 2: Nonrenormalization theorems}",
     journal        = "Nucl.Phys.",
     volume         = "B648",
     pages          = "3-34",
     doi            = "10.1016/S0550-3213(02)00979-3",
     year           = "2003",
     eprint         = "hep-th/0209064",
     archivePrefix  = "arXiv",
     primaryClass   = "hep-th",
     reportNumber   = "HU-EP-02-29, NUB-3231",
     SLACcitation   = "
}

@article{Dolan:2007eh,
     author         = "Dolan, Louise and Goddard, Peter",
     title          = "{Current Algebra on the Torus}",
     journal        = "Commun.Math.Phys.",
     volume         = "285",
     pages          = "219-264",
     doi            = "10.1007/s00220-008-0542-1",
     year           = "2009",
     eprint         = "0710.3743",
     archivePrefix  = "arXiv",
     primaryClass   = "hep-th",
     SLACcitation   = "
}

@incollection{ManinIterMod,
 title={Iterated integrals of modular forms and noncommutative modular symbols},
 author={Manin, Yuri I.},
 booktitle={Algebraic Geometry and Number Theory},
 pages={565--597},
 year={2006},
 publisher={Springer}
}

@article{BroadKrei,
 title={Association of multiple zeta values with positive knots via Feynman diagrams up to 9 loops},
 author={Broadhurst, David J and Kreimer, Dirk},
 journal={Physics Letters B},
 volume={393},
 number={3},
 pages={403--412},
 year={1997},
 publisher={Elsevier}
}

@incollection {GKZ,
   AUTHOR = {Gangl, Herbert and Kaneko, Masanobu and Zagier, Don},
    TITLE = {Double zeta values and modular forms},
BOOKTITLE = {Automorphic forms and zeta functions},
    PAGES = {71--106},
PUBLISHER = {World Sci. Publ., Hackensack, NJ},
     YEAR = {2006},
  MRCLASS = {11M41 (11F11)},
 MRNUMBER = {2208210 (2006m:11138)},
MRREVIEWER = {Hirofumi Tsumura},
      DOI = {10.1142/9789812774415_0004},
      URL = {http://dx.doi.org/10.1142/9789812774415\_0004},
}

@article {Rac,
   AUTHOR = {Racinet, Georges},
    TITLE = {Doubles m\'elanges des polylogarithmes multiples aux racines
             de l'unit\'e},
  JOURNAL = {Publ. Math. Inst. Hautes \'Etudes Sci.},
 FJOURNAL = {Publications Math\'ematiques. Institut de Hautes \'Etudes
             Scientifiques},
   VOLUME = { },
   NUMBER = {95},
     YEAR = {2002},
    PAGES = {185--231},
     ISSN = {0073-8301},
  MRCLASS = {11G55 (11M41)},
 MRNUMBER = {1953193 (2004c:11117)},
MRREVIEWER = {Jan Nekov{\'a}{\v{r}}},
      DOI = {10.1007/s102400200004},
      URL = {http://dx.doi.org/10.1007/s102400200004},
}

@incollection {Deligne,
   AUTHOR = {Deligne, P.},
    TITLE = {Le groupe fondamental de la droite projective moins trois
             points},
BOOKTITLE = {Galois groups over ${\bf Q}$ ({B}erkeley, {CA}, 1987)},
   SERIES = {Math. Sci. Res. Inst. Publ.},
   VOLUME = {16},
    PAGES = {79--297},
PUBLISHER = {Springer, New York},
     YEAR = {1989},
  MRCLASS = {14G25 (11G35 11M06 11R70 14F35 19E99 19F27)},
 MRNUMBER = {1012168 (90m:14016)},
MRREVIEWER = {James Milne},
      DOI = {10.1007/978-1-4613-9649-9_3},
      URL = {http://dx.doi.org/10.1007/978-1-4613-9649-9\_3},
}

@article{Sogaard:2014jla,
     author         = "S{\o}gaard, Mads and Zhang, Yang",
     title          = "{Elliptic Functions and Maximal Unitarity}",
     journal        = "Phys.Rev.",
     number         = "8",
     volume         = "D91",
     pages          = "081701",
     doi            = "10.1103/PhysRevD.91.081701",
     year           = "2015",
     eprint         = "1412.5577",
     archivePrefix  = "arXiv",
     primaryClass   = "hep-th",
     SLACcitation   = "
}

@article{BauSch,
     author         = "S. Baumard and L. Schneps",
     title          = "{Relations dans l'alg\`{e}bre de Lie fondamentale des motifs elliptiques mixtes}",
     year           = "2013",
     eprint         = "1310.5833",
     archivePrefix  = "arXiv",
     primaryClass   = "math",
}

@article{Brown:1504,
     author         = "F. Brown",
     title          = "{Zeta elements in depth 3 and the fundamental Lie algebra of a punctured elliptic curve}",
     year           = "2015",
     eprint         = "1504.04737",
     archivePrefix  = "arXiv",
     primaryClass   = "math",
}

@misc{Brown:LetterToMatthes,
     author         = "F. Brown",
     title          = "{Letter to Nils Matthes}",
     year           = "2015",
}

@article{Lipatov:1976zz,
      author         = "Lipatov, L. N.",
      title          = "{Reggeization of the Vector Meson and the Vacuum
                        Singularity in Nonabelian Gauge Theories}",
      journal        = "Sov. J. Nucl. Phys.",
      volume         = "23",
      year           = "1976",
      pages          = "338-345",
      cnote           = "[Yad. Fiz.23,642(1976)]",
      SLACcitation   = "
}

@article{Fadin:1975cb,
      author         = "Fadin, Victor S. and Kuraev, E. A. and Lipatov, L. N.",
      title          = "{On the Pomeranchuk Singularity in Asymptotically Free
                        Theories}",
      journal        = "Phys. Lett.",
      volume         = "B60",
      year           = "1975",
      pages          = "50-52",
      doi            = "10.1016/0370-2693(75)90524-9",
      SLACcitation   = "
}

@article{Kuraev:1976ge,
      author         = "Kuraev, E. A. and Lipatov, L. N. and Fadin, Victor S.",
      title          = "{Multi - Reggeon Processes in the Yang-Mills Theory}",
      journal        = "Sov. Phys. JETP",
      volume         = "44",
      year           = "1976",
      pages          = "443-450",
      cnote           = "[Zh. Eksp. Teor. Fiz.71,840(1976)]",
      SLACcitation   = "
}

@article{Balitsky:1978ic,
      author         = "Balitsky, I. I. and Lipatov, L. N.",
      title          = "{The Pomeranchuk Singularity in Quantum Chromodynamics}",
      journal        = "Sov. J. Nucl. Phys.",
      volume         = "28",
      year           = "1978",
      pages          = "822-829",
      cnote           = "[Yad. Fiz.28,1597(1978)]",
      SLACcitation   = "
}

@article{Bartels:2008sc,
      author         = "Bartels, J. and Lipatov, L. N. and Sabio Vera, Agustin",
      title          = "{N=4 supersymmetric Yang Mills scattering amplitudes at
                        high energies: The Regge cut contribution}",
      journal        = "Eur. Phys. J.",
      volume         = "C65",
      year           = "2010",
      pages          = "587-605",
      doi            = "10.1140/epjc/s10052-009-1218-5",
      eprint         = "0807.0894",
      archivePrefix  = "arXiv",
      primaryClass   = "hep-th",
      reportNumber   = "CERN-PH-TH-2008-125, DESY-08-073",
      SLACcitation   = "
}

@article{Bartels:2008ce,
      author         = "Bartels, J. and Lipatov, L. N. and Sabio Vera, Agustin",
      title          = "{BFKL Pomeron, Reggeized gluons and Bern-Dixon-Smirnov
                        amplitudes}",
      journal        = "Phys. Rev.",
      volume         = "D80",
      year           = "2009",
      pages          = "045002",
      doi            = "10.1103/PhysRevD.80.045002",
      eprint         = "0802.2065",
      archivePrefix  = "arXiv",
      primaryClass   = "hep-th",
      reportNumber   = "DESY-08-015, CERN-PH-TH-2008-027",
      SLACcitation   = "
}

@article{DelDuca:2009au,
      author         = "Del Duca, Vittorio and Duhr, Claude and Smirnov, Vladimir
                        A.",
      title          = "{An Analytic Result for the Two-Loop Hexagon Wilson Loop
                        in N = 4 SYM}",
      journal        = "JHEP",
      volume         = "03",
      year           = "2010",
      pages          = "099",
      doi            = "10.1007/JHEP03(2010)099",
      eprint         = "0911.5332",
      archivePrefix  = "arXiv",
      primaryClass   = "hep-ph",
      reportNumber   = "IPPP-09-92, DCPT-09-184",
      SLACcitation   = "
}

@article{DelDuca:2010zg,
      author         = "Del Duca, Vittorio and Duhr, Claude and Smirnov, Vladimir
                        A.",
      title          = "{The Two-Loop Hexagon Wilson Loop in N = 4 SYM}",
      journal        = "JHEP",
      volume         = "05",
      year           = "2010",
      pages          = "084",
      doi            = "10.1007/JHEP05(2010)084",
      eprint         = "1003.1702",
      archivePrefix  = "arXiv",
      primaryClass   = "hep-th",
      reportNumber   = "IPPP-10-21, DCPT-10-42, CERN-PH-TH-2010-059",
      SLACcitation   = "
}

@article{Fadin:2011we,
      author         = "Fadin, V. S. and Lipatov, L. N.",
      title          = "{BFKL equation for the adjoint representation of the
                        gauge group in the next-to-leading approximation at N=4
                        SUSY}",
      journal        = "Phys. Lett.",
      volume         = "B706",
      year           = "2012",
      pages          = "470-476",
      doi            = "10.1016/j.physletb.2011.11.048",
      eprint         = "1111.0782",
      archivePrefix  = "arXiv",
      primaryClass   = "hep-th",
      reportNumber   = "DESY-11-191",
      SLACcitation   = "
}

@article{Faddeev:1994zg,
      author         = "Faddeev, L. D. and Korchemsky, G. P.",
      title          = "{High-energy QCD as a completely integrable model}",
      journal        = "Phys. Lett.",
      volume         = "B342",
      year           = "1995",
      pages          = "311-322",
      doi            = "10.1016/0370-2693(94)01363-H",
      eprint         = "hep-th/9404173",
      archivePrefix  = "arXiv",
      primaryClass   = "hep-th",
      reportNumber   = "ITP-SB-94-14",
      SLACcitation   = "
}

@article{Lipatov:1993yb,
      author         = "Lipatov, L. N.",
      title          = "{High-energy asymptotics of multicolor QCD and exactly
                        solvable lattice models}",
      year           = "1993",
      eprint         = "hep-th/9311037",
      archivePrefix  = "arXiv",
      primaryClass   = "hep-th",
      reportNumber   = "DFPD-93-TH-70",
      SLACcitation   = "
}

@article{Lipatov:1994xy,
      author         = "Lipatov, L. N.",
      title          = "{Asymptotic behavior of multicolor QCD at high energies
                        in connection with exactly solvable spin models}",
      journal        = "JETP Lett.",
      volume         = "59",
      year           = "1994",
      pages          = "596-599",
      cnote           = "[Pisma Zh. Eksp. Teor. Fiz.59,571(1994)]",
      SLACcitation   = "
}

@article{Prygarin:2011gd,
      author         = "Prygarin, Alexander and Spradlin, Marcus and Vergu,
                        Cristian and Volovich, Anastasia",
      title          = "{All Two-Loop MHV Amplitudes in Multi-Regge Kinematics
                        From Applied Symbology}",
      journal        = "Phys. Rev.",
      volume         = "D85",
      year           = "2012",
      pages          = "085019",
      doi            = "10.1103/PhysRevD.85.085019",
      eprint         = "1112.6365",
      archivePrefix  = "arXiv",
      primaryClass   = "hep-th",
      reportNumber   = "BROWN-HET-1624",
      SLACcitation   = "
}

@article{Bartels:2011ge,
      author         = "Bartels, J. and Kormilitzin, A. and Lipatov, L. N. and
                        Prygarin, A.",
      title          = "{BFKL approach and $2 \to 5$ maximally helicity violating
                        amplitude in ${\cal N}=4$ super-Yang-Mills theory}",
      journal        = "Phys. Rev.",
      volume         = "D86",
      year           = "2012",
      pages          = "065026",
      doi            = "10.1103/PhysRevD.86.065026",
      eprint         = "1112.6366",
      archivePrefix  = "arXiv",
      primaryClass   = "hep-th",
      SLACcitation   = "
}

@article{Bartels:2013jna,
      author         = "Bartels, Jochen and Kormilitzin, Andrey and Lipatov, Lev",
      title          = "{Analytic structure of the $n=7$ scattering amplitude in
                        $\mathcal{N}=4$ SYM theory in the multi-Regge kinematics:
                        Conformal Regge pole contribution}",
      journal        = "Phys. Rev.",
      volume         = "D89",
      year           = "2014",
      number         = "6",
      pages          = "065002",
      doi            = "10.1103/PhysRevD.89.065002",
      eprint         = "1311.2061",
      archivePrefix  = "arXiv",
      primaryClass   = "hep-th",
      reportNumber   = "DESY-13-209",
      SLACcitation   = "
}

@article{Bartels:2014jya,
      author         = "Bartels, Jochen and Kormilitzin, Andrey and Lipatov, Lev
                        N.",
      title          = "{Analytic structure of the $n=7$ scattering amplitude in
                        $\mathcal{N}=4$ theory in multi-Regge kinematics:
                        Conformal Regge cut contribution}",
      journal        = "Phys. Rev.",
      volume         = "D91",
      year           = "2015",
      number         = "4",
      pages          = "045005",
      doi            = "10.1103/PhysRevD.91.045005",
      eprint         = "1411.2294",
      archivePrefix  = "arXiv",
      primaryClass   = "hep-th",
      reportNumber   = "DESY-14-171",
      SLACcitation   = "
}

@article{Bartels:2010ej,
      author         = "Bartels, J. and Kotanski, J. and Schomerus, V.",
      title          = "{Excited Hexagon Wilson Loops for Strongly Coupled N=4
                        SYM}",
      journal        = "JHEP",
      volume         = "01",
      year           = "2011",
      pages          = "096",
      doi            = "10.1007/JHEP01(2011)096",
      eprint         = "1009.3938",
      archivePrefix  = "arXiv",
      primaryClass   = "hep-th",
      reportNumber   = "DESY-10-157",
      SLACcitation   = "
}

@article{Bartels:2012gq,
      author         = "Bartels, J. and Schomerus, V. and Sprenger, M.",
      title          = "{Multi-Regge Limit of the n-Gluon Bubble Ansatz}",
      journal        = "JHEP",
      volume         = "11",
      year           = "2012",
      pages          = "145",
      doi            = "10.1007/JHEP11(2012)145",
      eprint         = "1207.4204",
      archivePrefix  = "arXiv",
      primaryClass   = "hep-th",
      reportNumber   = "DESY-12-122",
      SLACcitation   = "
}

@article{Bartels:2013dja,
      author         = "Bartels, J. and Kotanski, J. and Schomerus, V. and
                        Sprenger, M.",
      title          = "{The Excited Hexagon Reloaded}",
      year           = "2013",
      eprint         = "1311.1512",
      archivePrefix  = "arXiv",
      primaryClass   = "hep-th",
      reportNumber   = "DESY-13-197",
      SLACcitation   = "
}

@article{Bartels:2014ppa,
      author         = "Bartels, J. and Schomerus, V. and Sprenger, M.",
      title          = "{Heptagon Amplitude in the Multi-Regge Regime}",
      journal        = "JHEP",
      volume         = "10",
      year           = "2014",
      pages          = "67",
      doi            = "10.1007/JHEP10(2014)067",
      eprint         = "1405.3658",
      archivePrefix  = "arXiv",
      primaryClass   = "hep-th",
      reportNumber   = "DESY-14-076",
      SLACcitation   = "
}

@article{Bartels:2014mka,
      author         = "Bartels, J. and Schomerus, V. and Sprenger, M.",
      title          = "{The Bethe roots of Regge cuts in strongly coupled $
                        \mathcal{N}=4 $ SYM theory}",
      journal        = "JHEP",
      volume         = "07",
      year           = "2015",
      pages          = "098",
      doi            = "10.1007/JHEP07(2015)098",
      eprint         = "1411.2594",
      archivePrefix  = "arXiv",
      primaryClass   = "hep-th",
      reportNumber   = "DESY-14-208",
      SLACcitation   = "
}

@article{Basso:2013aha,
      author         = "Basso, Benjamin and Sever, Amit and Vieira, Pedro",
      title          = "{Space-time S-matrix and Flux tube S-matrix II.
                        Extracting and Matching Data}",
      journal        = "JHEP",
      volume         = "01",
      year           = "2014",
      pages          = "008",
      doi            = "10.1007/JHEP01(2014)008",
      eprint         = "1306.2058",
      archivePrefix  = "arXiv",
      primaryClass   = "hep-th",
      SLACcitation   = "
}

@article{Basso:2014koa,
      author         = "Basso, Benjamin and Sever, Amit and Vieira, Pedro",
      title          = "{Space-time S-matrix and Flux-tube S-matrix III. The
                        two-particle contributions}",
      journal        = "JHEP",
      volume         = "08",
      year           = "2014",
      pages          = "085",
      doi            = "10.1007/JHEP08(2014)085",
      eprint         = "1402.3307",
      archivePrefix  = "arXiv",
      primaryClass   = "hep-th",
      SLACcitation   = "
}

@article{Basso:2014nra,
      author         = "Basso, Benjamin and Sever, Amit and Vieira, Pedro",
      title          = "{Space-time S-matrix and Flux-tube S-matrix IV. Gluons
                        and Fusion}",
      journal        = "JHEP",
      volume         = "09",
      year           = "2014",
      pages          = "149",
      doi            = "10.1007/JHEP09(2014)149",
      eprint         = "1407.1736",
      archivePrefix  = "arXiv",
      primaryClass   = "hep-th",
      SLACcitation   = "
}

@article{Basso:2013vsa,
      author         = "Basso, Benjamin and Sever, Amit and Vieira, Pedro",
      title          = "{Spacetime and Flux Tube S-Matrices at Finite Coupling
                        for N=4 Supersymmetric Yang-Mills Theory}",
      journal        = "Phys. Rev. Lett.",
      volume         = "111",
      year           = "2013",
      number         = "9",
      pages          = "091602",
      doi            = "10.1103/PhysRevLett.111.091602",
      eprint         = "1303.1396",
      archivePrefix  = "arXiv",
      primaryClass   = "hep-th",
      SLACcitation   = "
}

@article{Drummond:2013nda,
      author         = "Drummond, James and Duhr, Claude and Eden, Burkhard and
                        Heslop, Paul and Pennington, Jeffrey and Smirnov, Vladimir
                        A.",
      title          = "{Leading singularities and off-shell conformal
                        integrals}",
      journal        = "JHEP",
      volume         = "08",
      year           = "2013",
      pages          = "133",
      doi            = "10.1007/JHEP08(2013)133",
      eprint         = "1303.6909",
      archivePrefix  = "arXiv",
      primaryClass   = "hep-th",
      reportNumber   = "HU-EP-13-15, IPPP-13-09, DCPT-13-18, --SLAC-PUB-15409,
                        LAPTH-016-13, CERN-PH-TH-2013-058, HU-MATHEMATIK:2013-06",
      SLACcitation   = "
}

@article{Lipatov:2010ad,
      author         = "Lipatov, L. N. and Prygarin, A.",
      title          = "{BFKL approach and six-particle MHV amplitude in N=4
                        super Yang-Mills}",
      journal        = "Phys. Rev.",
      volume         = "D83",
      year           = "2011",
      pages          = "125001",
      doi            = "10.1103/PhysRevD.83.125001",
      eprint         = "1011.2673",
      archivePrefix  = "arXiv",
      primaryClass   = "hep-th",
      reportNumber   = "DESY-10-173",
      SLACcitation   = "
}

@article{Bartels:2011xy,
      author         = "Bartels, J. and Lipatov, L. N. and Prygarin, A.",
      title          = "{Collinear and Regge behavior of 2 $\rightarrow$ 4 MHV amplitude in
                        N = 4 super Yang-Mills theory}",
      year           = "2011",
      eprint         = "1104.4709",
      archivePrefix  = "arXiv",
      primaryClass   = "hep-th",
      reportNumber   = "DESY-11-052",
      SLACcitation   = "
}

@article{Bauer:2000cp,
      author         = "Bauer, Christian W. and Frink, Alexander and Kreckel,
                        Richard",
      title          = "{Introduction to the GiNaC framework for symbolic
                        computation within the C++ programming language}",
      journal        = "J. Symb. Comput.",
      volume         = "33",
      year           = "2000",
      pages          = "1",
      eprint         = "cs/0004015",
      archivePrefix  = "arXiv",
      primaryClass   = "cs-sc",
      reportNumber   = "MZ-TH-00-17",
      SLACcitation   = "
}

@inproceedings{Papathanasiou:2014yva,
      author         = "Papathanasiou, Georgios",
      title          = "{Evaluating the six-point remainder function near the
                        collinear limit}",
      crossref       = "Rencontres",
      journal        = "Int. J. Mod. Phys.",
      volume         = "A29",
      year           = "2014",
      number         = "27",
      doi            = "10.1142/S0217751X14501541",
      eprint         = "1406.1123",
      archivePrefix  = "arXiv",
      primaryClass   = "hep-th",
      reportNumber   = "LAPTH-041-14",
      SLACcitation   = "
}

@proceedings{Rencontres,
     editor         = "{E. Aug\'{e}, J. Dumarchez, J. T. T. V\^{a}n}",
     booktitle      = "Proceedings, 49th Rencontres de Moriond on QCD and High
                        Energy Interactions",
     publisher      = "ARISF",
     year           = "2014",
}

@article{Prygarin:2011gd,
      author         = "Prygarin, Alexander and Spradlin, Marcus and Vergu,
                        Cristian and Volovich, Anastasia",
      title          = "{All Two-Loop MHV Amplitudes in Multi-Regge Kinematics
                        From Applied Symbology}",
      journal        = "Phys. Rev.",
      volume         = "D85",
      year           = "2012",
      pages          = "085019",
      doi            = "10.1103/PhysRevD.85.085019",
      eprint         = "1112.6365",
      archivePrefix  = "arXiv",
      primaryClass   = "hep-th",
      reportNumber   = "BROWN-HET-1624",
      SLACcitation   = "
}

\end{bibtex}

\bibliographystyle{nb}
\bibliography{MRK7}

\end{document}